\documentclass[journal]{IEEEtran}
\usepackage{cite}
\usepackage{amsmath} 

\usepackage{subfigure}
\ifCLASSINFOpdf
\usepackage[pdftex]{graphicx}
  \graphicspath{{../pdf/}{../jpeg/}}
  \DeclareGraphicsExtensions{.pdf,.jpeg,.png}
\else
  \usepackage[dvips]{graphicx}
  \graphicspath{{../eps/}}
  \DeclareGraphicsExtensions{.eps}
\fi  
\usepackage{cases}
\usepackage{stfloats}
\usepackage{amsfonts}
\usepackage{subeqnarray}
\usepackage{longtable}
\usepackage{supertabular}
\usepackage{setspace}
\usepackage{multirow}
\usepackage{booktabs}
\usepackage[ruled,linesnumbered]{algorithm2e}
\makeatletter
\newcommand{\nosemic}{\renewcommand{\@endalgocfline}{\relax}}
\newcommand{\dosemic}{\renewcommand{\@endalgocfline}{\algocf@endline}}
\let\oldnl\nl
\newcommand{\nonl}{\renewcommand{\nl}{\let\nl\oldnl}}
\makeatother
\usepackage[export]{adjustbox} 
\usepackage{booktabs}
\usepackage{setspace}
\usepackage{xcolor}
\usepackage{mathrsfs}
\usepackage{array}
\usepackage{amssymb}
\usepackage{amsthm}
\usepackage{microtype}
\usepackage{url}
\usepackage{amsfonts,amssymb}
\usepackage{dsfont}
\usepackage{mathtools}
\usepackage{xcolor,colortbl}
\usepackage{colortbl}
\usepackage{graphicx}
\usepackage[explicit]{titlesec}
\titlespacing{\section}{0pt}{1.2ex plus .0ex minus .0ex}{.3ex plus .0ex}
\titlespacing{\subsection}{0pt}{1.2ex plus .0ex minus .0ex}{.3ex plus .0ex}

\definecolor{Gray}{gray}{0.85}
\definecolor{Whitecolor}{rgb}{1,1,1}
\definecolor{cincinnati-red}{RGB}{190,0,0}
\definecolor{purpple}{RGB}{251,142,50}

\newcolumntype{a}{>{\columncolor{Gray}}c}

\allowdisplaybreaks
\begin{document}
\setstretch{0.93}
\title{\textls[-25]{A Data-driven Under Frequency Load Shedding Scheme in Power Systems}}
\author{Qianni~Cao,~\IEEEmembership{Student~Member,~IEEE},
Chen~Shen,~\IEEEmembership{Senior~Member,~IEEE}

}
        

\maketitle

\begin{abstract}
Under frequency load shedding (UFLS) serves as  the very last resort for preventing total blackouts and cascading events. Fluctuating operating conditions and weak resilience of the future grid require UFLS strategies adapt to various operating conditions and non-envisioned faults. This paper develops a novel data-enabled Koopman-based Load Shedding Scheme (KLS) to achieve the optimal one-shot load shedding for power system frequency safety. 
Our approach yields a network that facilitates a coordinate transformation from the delay embedded space to a new space, wherein the dynamics can be expressed in a linear manner. The network is specifically tailored to effectively track parameter variations in the dynamic model of the system.
To address approximation inaccuracies and the discrete nature of load shedding, a safety margin tuning scheme is integrated into the KLS framework, ensuring that the system frequency trajectory remains within the safety range. Simulation results show the adaptability, prediction capability and control effect of the proposed UFLS strategy. 

\end{abstract}
\IEEEpeerreviewmaketitle




\section*{Nomenclature}
\addcontentsline{toc}{section}{Nomenclature}

\subsection*{Koopman observables} 
\begin{IEEEdescription}[\IEEEusemathlabelsep\IEEEsetlabelwidth{$aaaaaaaa$}]
	\item[$\boldsymbol{g}$]		 A set of finite Koopman observables which forms a subspace of the infinite dimensional Koopman observables.
	\item[$\bar{\boldsymbol{g}}$]	Finite Koopman observables identified from data.
	\item[$\hat{\boldsymbol{g}}$]	The trajectory of Koopman observables predicted by a Koopman linear representation obtained from data.
\end{IEEEdescription}

\section{Introduction}
\subsection{Motivation}
\IEEEPARstart{U}{nder} frequency events in bulk power systems are generally caused by sudden large active power deficits, such as generator tripping or load surges. Restraining such frequency unsafty is essential for the secure operation of power systems. 

Under frequency load shedding (UFLS) serves as the very last resort for preventing total blackouts and cascading events. The most standard and widely used strategy is to shed pre-assigned loads in discrete steps or stages. Each stage is implemented when the frequency is lower than a preset level of load shedding. 
In modern power systems, the uncertainty of renewable generation lead to ever-fluctuating operating conditions. The conventional UFLS strategy, which determines the load shedding amount based on anticipated operating conditions and faults, faces challenges in effectively working well under various operating conditions\cite{8850069}. Therefore, it is necessary to develop UFLS schemes that adapt to various operating conditions and events, decide the optimal load shedding amount to promise the hard limits on the frequency trajectories.

In this paper, an online data-enabled UFLS scheme is designed for emergency frequency control. With the frequency prediction capability under various operating conditions and power imbalances, it leverages online input/output measurements to achieves safe and optimal online control with minimal one-shot load shedding, instead of shedding loads at multi stages, thereby speeding up the recovery of system frequency.

\subsection{Literature Review}

There have been abundant works on frequency control. A challenging problem is optimal frequency trajectory tracking, where a control policy should drive the system frequency within safety range while minimizing load shedding amount. A general solution is to formulate the control strategy design problem as an optimization problem. The key ingredient for optimal control problem is an accurate state space model of the system. 

A standard solution is utilizing classical representations of system dynamics, such as the swing equation and the first-order Primary Frequency Response (PFR) dynamics. These methods require prior knowledge of system parameters\cite{8241866,6164299,8552409}, or need to know the power deficit in the system when faults occur \cite{Xu2015}. However, parameters such as inertia have become difficult to obtain accurately due to the high penetration of power electronics converter-interfaced devices \cite{9721402}. Moreover, the power deficit in the system cannot be measured directly in practical power networks. Although data-driven approaches are used to estimate these parameters\cite{9767658}, the accuracy of the model itself may be limited. For example, the traditional PFR model does not explicitly consider the impact of load-frequency dependence \cite{8336919}.

As power systems are becoming more complex and data is becoming more readily available, it is in favor to develop data-driven methods that use only input/output data measured from the unknown system \cite{HOU20133}. Data-driven methods are suitable for applications when models are too complex for control design, and when thorough modelling and parameter identification is too costly. In recent years, many studies have focused on the design of data-driven load shedding schemes for power systems. Ref.\cite{9626446} proposed a data-driven distributional soft actor critic (DSAC) method to solve the emergency frequency control problem. Ref. \cite{9779512} investigated the optimal frequency control problem using reinforcement learning with stability guarantees.

Fluctuating operating conditions and weak resilience of the future grid require frequency control strategies adapt to various operating conditions and non-envisioned faults. Consequently, data-driven control should possess the capability to accommodate various operating conditions and non-envisioned faults. However, for state-of-the-art data-driven UFLS, system responses are collected offline before the online control operation begins. These responses are used to estimate a model that matches the observed data in an appropriate sense. Nevertheless, offline observations are unable to cover all potential operating conditions and faults. Moreover, there is no guarantee that a system model trained for specific predetermined scenarios generalize to data outside of the distribution of the training set \cite{doi:10.1073/pnas.1906995116, 9626446, 9779512}. In Ref.\cite{9626446}, when addressing unforeseen operating conditions, the trained DSAC agent requires additional episodes to converge for strategy implementation. In Ref.\cite{9779512}, the reinforcement learning-based controller relies on dataset diversity to adapt to various system operating conditions and faults. Consequently, its control effect, evaluated by the frequency nadir and amount of control, degrades as the number of training trajectories decreases.
Nevertheless, ensuring that data-driven control strategies adapt to unanticipated operating conditions and faults is of importance. When incorporating an offline-trained system representation for online control, it should be able to track parameter variations in the system model using measurements.


Given that nonlinear system dynamics renders the optimal control problem intractable, it is desirable that the learned system model linear. The Koopman-based control framework \cite{KORDA2018297} has emerged as a dominant perspective in data-driven control due to its ability to provide linear representations of system dynamics in the context of black-box systems. Finite Koopman invariant subspace is often approximated to gear toward prediction and control since Koopman operator is infinite-dimensional. Dynamic Mode Decomposition (DMD) and Extended DMD (EDMD) stand as leading approaches for approximating the Koopman operator. However, they rely on prior knowledge of the underlying nonlinear dynamics to manually select suitable observation functions. Some approaches propose utilizing deep neural networks with sufficient layers to derive more effective and accurate observable functions from measurement data \cite{ping2021deep}. Nonetheless, all these methods rely on historical data to obtain fixed Koopman linear representations. When the system's operating conditions change or different faults occur, the parameters in the dynamical model of the system change. Adapting Koopman linear representations to parameter variations in the system model remains an open challenge.

Moreover, for the state-of-the-art Koopman operator approximation methods, representation errors of Koopman operators are inevitable \cite{doi:10.1137/21M1401243,Kaiser_2021}. Inaccurate estimations and predictions and may degrade the quality of the obtained optimal control. In Ref. \cite{10135149}, the impact of representation errors of Koopman eigenpairs on the control performance of LQR controllers was evaluated. The adaptive nature of closed-loop control theoretically allows one to compensate for modeling discrepancies and to account for disturbances\cite{doi:10.1137/21M1401243}. Nevertheless, in practice, load shedding measures are rarely implemented as closed-loop strategies \cite{Shen_Cao_Jia_Chen_Huang_2022}. Hence, for Koopman-based UFLS, it is necessary to address the potential impact of representation errors on open-loop control.
 
In most load shedding schemes, it is often assumed that the load shedding amount at each bus can be a continuous value\cite{9540244,9626446}. This assumption overlooks the fact that load shedding is accomplished by shedding candidate feeders. In practical engineering, the feasible amount of load shedding is restricted to discrete values, with each discrete interval representing the load associated with a feeder.

Although the optimal control problem can be formulated as an MILP problem to decide whether to shed a feeder or not \cite{MAYNE20142967}, solving MILP is an NP-hard problem, making it difficult for online control. Therefore, Ref.\cite{9424434,AZUMA2008396,1098527} decides the control value at a stage beforehand and then rounds it to the nearest discrete value. However, this leads to a discrepancy between the actual and optimal shedding amounts, which affects the dynamic behavior of the system after control. Consequently, when designing the control strategy, it is essential to explicitly consider the potential effects of the discrepancy and ensure that the dynamic system remains within the safety range.

\subsection{Contribution}

According to the literature review, to address the frequency stability issue in modern power grids with high penetration of renewable energy sources, we face three main challenges: 1) system identification for optimal control under unanticipated operating conditions and power imbalance; 2) analyzing the system performance when applying control policies computed from an inaccurate Koopman representations; 3) designing a control strategy that works with feasible discrete load shedding values.

In response to the aforementioned research gaps, the main contributions of this study are outlined as follows:

\begin{enumerate}
    \item We introduces a novel data-enabled predictive control algorithm, referred to as KLS, that achieves optimal one-shot load shedding for power system frequency safety. The KLS algorithm demonstrates adaptability to diverse operating conditions and under frequency events, allowing for precise load shedding strategies.
    
    \item We investigate how approximation inaccuracies in the Koopman linear representations influence the control strategies and controlled frequency trajectories. 
    
    \item By formulating a safety margin tuning scheme within the framework of KLS, we ensure that the system frequency trajectory remains within the prescribed hard limits when approximation inaccuracies exist and the feasible amount of load shedding is restricted to discrete values.
\end{enumerate}

The rest of this paper is organized as follows. Section II proposes the KLS strategy for under frequency events. Section III investigates how approximation inaccuracies in the Koopman linear representations influence the control effect and a safety margin tuning method is proposed. In Section IV, an CIGRE-LF system case is presented and the effectiveness of the proposed control strategy is verified. Section V provides the conclusion.


\section{Emergency Frequency Control Design}
\label{Emergency_Frequency_Controller_Design}

In this section, we design a deep neural network to learn a coordinate transformation from the delay embedded measurement space into a new space where it is possible to represent the frequency dynamics linearly. An optimal control problem is then formulated to obtain the one-shot load shedding amount.

Let an autonomous nonlinear dynamic system be governed by
\begin{align}
    {\boldsymbol{x}_{t+1}}=\boldsymbol{f}(\boldsymbol{x}_{t},\boldsymbol{y}_{t})\label{autonomous_nonlinear_dynamical_system}
\end{align}
where $t=1,2,...T$, $T$ is the prediction horizon, $\boldsymbol{x}\in {R}^{{n}_{x}}$ is the state, $\boldsymbol{y}\in {{R}^{{{n}_{y}}}}$ denotes algebraic variables and $\boldsymbol{f}(\cdot)$ is a nonlinear function. Considering the computational inefficiency of calculating optimal control for high-dimensional nonlinear dynamical functions, Koopman theory \cite{koopman1931hamiltonian} provides a perspective that nonlinear dynamics can be represented in terms of an infinite-dimensional linear operator acting on the space of all possible measurement functions of the system. Even if the function $\boldsymbol{f}(\cdot)$ is unknown, it is still possible to estimate the Koopman operator using the system's measurements.

However, the estimation of the Koopman operator relies exclusively on data, either numerical or experimental. In the context of power systems, it is common practice to rely on numerical data acquired from simulations. When collecting the dataset, it is necessary to preset the operation conditions and emergency events to trigger the system dynamics. Diverse operational conditions lead to variations in the parameters of the grid's state space model given in Eq.\eqref{autonomous_nonlinear_dynamical_system}, and also, the Koopman linear representations\cite{doi:10.1137/21M1401243}. Thus, a significant challenge in linear predictive system modeling is adaption to different operating conditions and faults in the training set.

Although it is feasible to incorporate a wide range of operation conditions and events within the training set, it is impractical to exhaustively account for every scenario. Therefore, the linear representation should also possess the ability to generalize, allowing the linear dynamic model of the system, which is trained for specific predetermined scenarios, to extend its prediction capability to systems operating under non-predefined conditions and faults that are not included in the sample set.

In order to explicitly represent the variations of operating conditions and the complexity of emergency events in the state space model of grids Eq. \eqref{autonomous_nonlinear_dynamical_system}, we utilize a vector of variables $\boldsymbol{m}$ to represent a subset of uncertain model parameters that are challenging to obtain online. To incorporate the uncertainty of these parameters into the system model, a modified model is evaluated as
\begin{align}
    {\boldsymbol{x}_{t+1}}=\boldsymbol{f}(\boldsymbol{x}_{t},\boldsymbol{y}_{t},\boldsymbol{m}_{t})  \notag \\ 
    {\boldsymbol{m}_{t+1}}=\boldsymbol{h}(\boldsymbol{x}_{t},\boldsymbol{y}_{t},\boldsymbol{m}_{t}) \label{a_modified_model}
\end{align}

Compared with Eq.\eqref{autonomous_nonlinear_dynamical_system}, Eq.\eqref{a_modified_model} provides a more general form of a deterministic power system model which accounts for uncertainties. In Eq.\eqref{a_modified_model}, $\boldsymbol{x}_{a}^{\top}=[\boldsymbol{x}^{\top}\ \boldsymbol{m}^{\top}]$ can be defined as pseudo-state variables. The augmented model was first introduced in Ref. \cite{9749019}.

\vspace{0.1cm}
\textit{\textbf{Remark 1}}: Many model parameters are typically assumed to be time-invariant, although their initial values may vary across different data samples, such as system inertia. There also exist time-varying parameters, but their temporal variations can be captured using discrete-time dynamic equations, as presented in Eq. \eqref{a_modified_model}, for instance, the power deficit in the system.

\vspace{0.2cm}
Since $\boldsymbol{m}$ are hard to measure, which constitutes hidden, or latent variables that are not directly measured but are dynamically important. Thus, the challenge of adapting the linear representations to accommodate the parameters variations transforms into the challenge of accounting for the hidden variable in the model.

Time-delay embedding provides an approach to augment these hidden variables, and under certain conditions, given by Takens’ embedding theorem \cite{Takens_1981}, the delay-augmented state yields an attractor that is diffeomorphic to the underlying, though unmeasured, full-state attractor \cite{bakarji2022discovering}. Here, we design a deep neural network to learn a coordinate transformation from the delay embedded space into a new space. In the new space, it is possible to represent the dynamics in a linear form. Additionally, the network also tracks the parameter variations in the system with input/output data.

The deep neural network is illustrated in Appendix \ref{Network Architecture and Training}. The data set collected for training the network is described as follows. For a given power system, a specific anticipated operating condition $\alpha\in\mathfrak{A}$ is considered, and a representative fault $\beta\in\mathfrak{B}$ is introduced. $\mathfrak{A}$ represents a predefined set of typical operating conditions. $\mathfrak{B}$ denotes a predefined set of typical faults. Additionally, load shedding amount $\boldsymbol{u}=[{{u}_{1}},...{{u}_{i}},...,{{u}_{I}}] \in \mathcal{U}$ are defined at each load node $i$, where ${{u}_{i}}(i=1,2,...I)$ (in per unit, where the load level at bus $i$ is the base value for ${u}_{i}$) is a uniformly distributed random number between 0 and 1. Subsequently, time-series data of the system's inertia center frequency (referred to as the system frequency hereafter) $\Omega=\{{\omega}^{\alpha,\beta}|\alpha\in\mathfrak{A},\beta\in\mathfrak{B}\}$ are collected at time points $t=1,2,...,T$, resulting in a sequence of data $[\omega _{1}^{\alpha ,\beta },\omega _{2}^{\alpha ,\beta },...,\omega _{T}^{\alpha ,\beta }]$. The obtained $\Omega$ and $\boldsymbol{u}\in\mathcal{U}$ are utilized as training data for the linear prediction model. 

Details for the network architecture and the loss function are given in Appendix \ref{section_A}. The latent extraction layers in the network are specifically designed to monitor variations in the parameters of the state-space model.

Based on Koopman theory, we assume that the frequency dynamics of the system is governed by the linear dynamic system equation represented in Eq.\eqref{eq_linear_dynamic_system_equation}.
\begin{align}
    \boldsymbol{g}_{t+1}=\boldsymbol{A}{\boldsymbol{g}_{t}}+\boldsymbol{Bu}_{t}\label{eq_linear_dynamic_system_equation}
\end{align}
where 
\begin{align}
    \boldsymbol{g}_{t}=\left[ \begin{matrix}
   \omega_{t}  \\
   \boldsymbol{\varphi} ({\omega }_{t-\tau :t},\boldsymbol{y}_{t-\tau :t})  \\
\end{matrix} \right] \label{time_delay}
\end{align}
where $\omega_t$ is the deviation of frequency from its nominal value (in per unit) at time $t$, ${{\omega }_{t-\tau :t}}=[{{\omega }_{t-\tau }},{{\omega }_{t-\tau +\Delta t}},...,{{\omega }_{t}}],{{y}_{t-\tau :t}}=[{{y}_{t-\tau }},{{y}_{t-\tau +\Delta t}},...,{{y}_{t}}]$ represent the time series of system frequency (state variable) and voltages (algebraic variables), respectively. $\boldsymbol{\varphi}$ denotes a neural network with a prescribed activation function and connectivity structure. $\boldsymbol{g}$ is a set of finite Koopman observables which forms a subspace of the infinite dimensional Koopman observables. With the loss function and the algorithm provided in Appendix \ref{section_A} , it is feasible to approximate the parameters of $\boldsymbol{\varphi}$, along with the matrices $\boldsymbol{A}$ and $\boldsymbol{B}$.

After the parameters of $\boldsymbol{\varphi}$, $\boldsymbol{A}$, and $\boldsymbol{B}$ have been approximated from data, one can utilize Eq. \eqref{eq_linear_dynamic_system_equation} to predict the future trajectory of the system frequency variation, provided the control sequence $\boldsymbol{u}_t$, system frequency at the first time step $\omega_{1}^{\alpha,\beta}$, and a time series of ${\omega }_{1-\tau :1}^{\alpha,\beta}$ and $\boldsymbol{y}_{1-\tau :1}^{\alpha,\beta}$. Herein, we refer to the dynamic system described by Eq. (3) as a Koopman linear system.

\vspace{0.1cm}
\textit{\textbf{Remark 2}}: The time intervals between the time points $t=1,2,...,T$ may not be consistent with the time intervals between $t-\tau,t-\tau+\Delta t,...,t$. In Section \ref{case_study}, the time intervals between $t=1,2,...,T$ are set to 1 s, while in $t-\tau,t-\tau+\Delta t,...,t$, $\Delta t$ is set to 1 ms.

\subsection{Koopman-Operator-Based Emergency Frequency Control Strategy} \label{Emergency_Frequency_Control_Strategy}

Combined with the Koopman model predictive control proposed in Ref. \cite{KORDA2018149}, the optimal shedding amount is obtained by solving the following optimal control problem. 
\begin{align}
& \min~~\ \ \ \boldsymbol{u}^{\text{T}}\boldsymbol{R}\boldsymbol{u}~ \notag \\ 
 & \text{s.t.}\ \ \ \  \bar{\omega}_{t}\ge{\omega}_{\text{min}},\ \ t=1,2,..., T\notag \\
 &\ \ \ \ \ \ \  \bar{\omega}_{T}\ge{\omega}_{\infty \text{min}} \notag \\
 &\ \ \ \ \ \ \  \boldsymbol{g}_{t+1}=\boldsymbol{A}{\boldsymbol{g}_{t}}+\boldsymbol{Bu}_{t},\ \ t=1,2,..., T \label{optimal_control_problem}
\end{align}
where $\boldsymbol{R}$ represents a positive definite matrix used to represent the cost of load shedding at each bus, $\bar{\omega}_{t}$ denotes system frequency at time $t$ predicted by Koopman linear system, ${\omega}_{\text{min}}$ is the minimal allowed system frequency, ${\omega}_{\infty \text{min}}$ is the minimum allowed steady-state frequency. Here, we assume that the prediction length $T$ is sufficiently long for the system frequency to reach a steady state by time $T$. The optimal control problem Eq.\eqref{optimal_control_problem} is a quadratic programming problem with $\boldsymbol{R}$ being a positive definite matrix. This problem can be solved in polynomial time. 

\vspace{0.1cm}
\textit{\textbf{Remark 3}}: The values of ${\omega}_{\text{min}}$ and ${\omega}_{\infty \text{min}}$ can be adjusted based on the interplay between data-driven load shedding and the traditional UFLS scheme. The traditional UFLS scheme, as it initiates load shedding after a certain deviation in system frequency occurs (e.g., when the system frequency drops to 49Hz), may lead to unexpected severe consequences due to the delayed timing of load shedding, such as greater power deficits and consequently higher load losses. If the objective of data-driven load shedding is to avoid triggering the traditional UFLS scheme, ${\omega}_{\text{min}}$ can be set to 49.0Hz. On the other hand, if the data-driven load shedding aims to fully replace the traditional UFLS scheme and ensure that the system's minimum frequency remains above the minimum operating frequency of synchronous generators (e.g., 47.0Hz), ${\omega}_{\text{min}}$ can be set to 47.0Hz. In Section \ref{case_study}, we choose ${\omega}_{\text{min}}$ as 49.0Hz as an illustrative example to show the effectiveness of KLS. 

In practice, continuous adjustment of load shedding value is difficult to achieve, and it is often necessary to choose whether or not to shed a load on a particular feeder line, resulting in a series of discrete values for the actual load shedding. Let the optimal load shedding value obtained by solving the optimal control problem Eq.\eqref{optimal_control_problem} be denoted as ${\boldsymbol{\bar{u}}_{*}}=[{{\bar{u}}_{1*}},...{{\bar{u}}_{i*}}...,{{\bar{u}}_{I*}}]$ the actual load shedding amount would then be given as
\begin{align}
    {{Q}_{d}}\left( {{{\bar{u}}}_{i*}} \right)=\left\{ \begin{matrix}
   nd & nd\le {{{\bar{u}}}_{i*}}<n\left( d+0.5 \right)  \\
   n\left( d+1 \right) & n\left( d+0.5 \right)\le {{{\bar{u}}}_{i*}}<n\left( d+1 \right)  \\
\end{matrix} \right. \label{actual_load_shedding_amount}
\end{align}
where $d$ represents the quantization interval, which physically refers to the load shedding amount on a single feeder line, $n$ represents a positive integer, and $Q_d({\bar{u}}_{i\ast})$ denotes the actual load shedding amount at each load node $i$ when the discrete interval is $d$. Solving the optimal control problem in Eq. \eqref{optimal_control_problem} and rounding the resulting solution as described in Eq. \eqref{actual_load_shedding_amount} is referred to as Koopman-based load shedding strategy (KLS).

\vspace{0.1cm}
\textit{\textbf{Remark 4}}: 
An alternative approach is to round the computed control quantity to the nearest larger value, as outlined below.
\begin{align}
    {{Q}_{d}}\left( {{{\bar{u}}}_{i*}} \right)=
   n\left( d+1 \right)\ \ \ \  nd< {{{\bar{u}}}_{i*}}\le n\left( d+1 \right) \label{actual_load_shedding_amount_larger}
\end{align}

However, the strategy in Eq.\eqref{actual_load_shedding_amount_larger} results in over-shedding. Although, given the same system linear representation, the strategy presented in Eq.\eqref{actual_load_shedding_amount_larger} with a larger load shedding amount compared to  Eq.\eqref{actual_load_shedding_amount} is more likely to ensure that the system frequency does not violate safety constraints. Nevertheless, when implementing the strategy described in Eq.\eqref{actual_load_shedding_amount}, it is also possible to ensure the safety of the system frequency by tuning a safety margin in the constraints of Eq.\eqref{optimal_control_problem}. The design of the safety margin in KLS will be presented in Section \ref{Safety_margin_tuning}. By employing the load shedding amount given in Eq.\eqref{actual_load_shedding_amount} along with the safety margin, we achieve a reduced level of load shedding amount compared to the strategy in Eq.\eqref{actual_load_shedding_amount_larger}.


\section{Error Estimation and Safety Margin Design}
In Section \ref{Emergency_Frequency_Controller_Design}, we employed constraints in the optimal control problem to ensure that the frequency predicted by the Koopman prediction model remain above the acceptable minimum values. However, in actual power systems, the optimal load shedding obtained from Eq. \eqref{optimal_control_problem} may cause the system frequency to violate the prescribed hard limits. The main reasons are as follows. 

First, the training of $\boldsymbol{g}$, $\boldsymbol{A}$ and $\boldsymbol{B}$ terminates when the loss function is less than a specified tolerance. Therefore, potential inadequate training leads to representation errors in $\boldsymbol{g}$, $\boldsymbol{A}$ and $\boldsymbol{B}$. Denote the finite Koopman observables and Koopman system matrix identified from data as $\bar{\boldsymbol{A}},\bar{\boldsymbol{B}}$ and $\bar{\boldsymbol{g}}$. The representation errors manifests as minor prediction errors $\bar{\boldsymbol{g}}_{t+1}-\bar{\boldsymbol{A}}\bar{\boldsymbol{g}}_t-\bar{\boldsymbol{B}}{\boldsymbol{u}}_t$. Second, the actual load shedding amount may deviate from the optimal load shedding amount obtained since it should be rounded to a feasible value.

Hence, to ensure that the frequency remain above the acceptable minimum values,
we propose adding a safety margin to the frequency limits in constraints of Eq. \eqref{optimal_control_problem}. An explicit approach is introduced for calculating the safety margin. By integrating the safety margin, the optimal control strategy derived from KLS effectively ensures that the system frequency complies with the prescribed hard limits in the actual system.
\subsection{Impact of Koopman representation errors on control effects}
\label{Impact_of_Koopman_representation_errors_on_control_effects}

This subsection aims to determine whether even slight deviations between the learnt and the accurate Koopman linear dynamics can compromise the desired system properties, potentially violating the imposed hard limits.


In subsection \ref{Emergency_Frequency_Control_Strategy}, the formulation of optimal control problem Eq.\eqref{optimal_control_problem} is based on the assumption that the identification of $\boldsymbol{A},\boldsymbol{B}$ and $\boldsymbol{g}$ are accurate. However, in real applications, due to the training error, only $\bar{\boldsymbol{A}},\bar{\boldsymbol{B}}$ and $\bar{\boldsymbol{g}}$ can be identified from data. 

When solving the optimal control problem in Eq.\eqref{optimal_control_problem}, we can only employ $\bar{\boldsymbol{A}},\bar{\boldsymbol{B}}$, and $\bar{\boldsymbol{g}}$ identified from data, assuming that the equality in the following equation holds.
\begin{align}
    \bar{\boldsymbol{g}}_{t+1}=\bar{\boldsymbol{A}}\bar{\boldsymbol{g}}_t+\bar{\boldsymbol{B}}{\boldsymbol{u}}_t\label{inaccurate_Koopman_representation_no_beta}
\end{align}

With $\bar{\boldsymbol{A}},\bar{\boldsymbol{B}}$, and $\bar{\boldsymbol{g}}$, one utilize Eq. \eqref{inaccurate_Koopman_representation_no_beta} to predict the future trajectory of Koopman observables given $\boldsymbol{x}_{1}$ and $\boldsymbol{u}$. We denote the trajectory of Koopman observables predicted by Eq. \eqref{inaccurate_Koopman_representation_no_beta} as $[\hat{\boldsymbol{g}}_{2},...,\hat{\boldsymbol{g}}_{T}]$.


In order to analyze how Koopman representation errors infuence the control strategy, we assume that $\bar{\boldsymbol{A}}$ and $\bar{\boldsymbol{B}}$ can be expressed as
\begin{align}
    \bar{\boldsymbol{A}}=\boldsymbol{A}+\boldsymbol{\Delta}_{\boldsymbol{A}}
\end{align}
\begin{align}
    \bar{\boldsymbol{B}}=\boldsymbol{B}+\boldsymbol{\Delta}_{\boldsymbol{B}}
\end{align}
where $\boldsymbol{\Delta}_{\boldsymbol{A}}$ and $\boldsymbol{\Delta}_{\boldsymbol{B}}$ are bounded in terms of the induced norm as
\begin{align}
\left\|\boldsymbol{\Delta}_{\boldsymbol{A}}\right\|\le\varepsilon_{\boldsymbol{A}},\left\|\boldsymbol{\Delta}_{\boldsymbol{B}}\right\|\le\varepsilon_{\boldsymbol{B}}
\end{align}
where $\boldsymbol{A},\bar{\boldsymbol{A}}\in\mathbb{R}^{p\times p}$ and $\boldsymbol{B},\bar{\boldsymbol{B}}\in\mathbb{R}^{p\times q}$.

$\bar{\boldsymbol{g}}$ can be expressed as
\begin{align}
     \bar{\boldsymbol{g}}_t=\boldsymbol{g}_t+\boldsymbol{\mathfrak{g}}_t \label{eq_bar_g}
\end{align}
where ${\mathfrak{g}}_t\in\mathbb{R}^{p}$ is the discrepancy to the accurate observables $\boldsymbol{g}_t\in\mathbb{R}^{p}$, and we assume that $\left\|\mathfrak{g}_t\right\|$ is very small compared with $\left\|\boldsymbol{g}_t\right\|$ for $t=1,2,...,T$.




Combining Eq.\eqref{eq_bar_g}, dynamics in Eq.\eqref{inaccurate_Koopman_representation_no_beta} can be transformed into $\boldsymbol{g}$ coordinates:
\begin{align}
    \boldsymbol{g}_{t+1}=\Bar{\boldsymbol{A}}\boldsymbol{g}_{t}+\Bar{\boldsymbol{B}}\boldsymbol{u}_{t}+\Bar{\boldsymbol{A}}{\mathfrak{g}}(\boldsymbol{x}_t)-\boldsymbol{{\mathfrak{g}}}(\boldsymbol{x}_{t+1})\label{eq:learnt_nonlinear_system}
\end{align}

Define
\begin{align}
\boldsymbol{\gamma}(\textbf{\textit{x}}_{t+1})=\Bar{\boldsymbol{A}}\boldsymbol{\mathfrak{g}}(\boldsymbol{x}_t)-\boldsymbol{\mathfrak{g}}(\boldsymbol{x}_{t+1})
\end{align}
and $\left\|\boldsymbol{\gamma}(\textbf{\textit{x}}_{t+1})\right\|$ satisfies
\begin{align}
\left\|\boldsymbol{\gamma}(\textbf{\textit{x}}_{t+1})\right\|\le\left\|\Bar{\boldsymbol{A}}\boldsymbol{\mathfrak{g}}(\boldsymbol{x}_t)\right\|+\left\|\boldsymbol{\mathfrak{g}}(\boldsymbol{x}_{t+1})\right\|
\end{align}

Since $\left\|\Bar{\boldsymbol{A}}\boldsymbol{\mathfrak{g}}_t\right\|$ and $\left\|\boldsymbol{\mathfrak{g}}_{t+1}\right\|$ are small, we assume the disturbance is norm-bounded by:
\begin{align}
    \boldsymbol{\gamma}_{t}=\{\boldsymbol{\gamma}\in\mathbb{R}^{p}|\left\|\boldsymbol{\gamma}\right\|\le\varepsilon_{\boldsymbol{\gamma}}\}
\end{align}


For dynamics in Eq.\eqref{a_modified_model} and a fixed horizon $T$, we define $\boldsymbol{\mathcal{G}}(\boldsymbol{x})$, $\textbf{u}$ and $\boldsymbol{\gamma}$ as the stacked states, inputs and disturbances up to time $T$, i.e.,
\begin{align}
    \boldsymbol{\mathcal{G}}^{\top}(\boldsymbol{x})=\begin{bmatrix}\boldsymbol{g}^{\top}(\boldsymbol{x}_0)&\boldsymbol{g}^{\top}(\boldsymbol{x}_1)&\cdots&\boldsymbol{g}^{\top}(\boldsymbol{x}_T)
    \end{bmatrix}
\end{align}
\begin{align}
    \textbf{u}^{\top}=\begin{bmatrix}\boldsymbol{u}_0^{\top}&\boldsymbol{u}_1^{\top}&\cdots&\boldsymbol{u}_T^{\top}
    \end{bmatrix}
\end{align}
\begin{align}
    \boldsymbol{\gamma}^{\top}=[\boldsymbol{x}_0^{\top}\ \boldsymbol{\gamma}_0^{\top}\ \cdots\ \boldsymbol{\gamma}_{T-1}^{\top}]
\end{align}
Note that
we embed $x_0$ as the first component of the disturbance
process. 

Based on System Level Synthesis (SLS) \cite{chen2023robust,ANDERSON2019364}, we have a direct optimization over system responses $\mathbf{\mathcal{T}}_{\boldsymbol{g}},\mathbf{\mathcal{T}}_\textbf{u}$ defined as
\begin{equation}
  \left[
  \begin{gathered}
  \boldsymbol{\mathcal{G}}^{\top}(\boldsymbol{x})\\
  \textbf{u}_{*}\\
  \end{gathered}
  \right]= \left[
  \begin{gathered}
 \mathbf{\mathcal{T}}_{\boldsymbol{g}}\\
  \mathbf{\mathcal{T}}_\textbf{u}
  \end{gathered}
  \right]\boldsymbol{\gamma}
\end{equation}
where $\mathbf{\mathcal{T}}_{\boldsymbol{g}}\in\mathcal{L}^{nT\times nT},\mathbf{\mathcal{T}}_{\textbf{u}}\in\mathcal{L}^{mT\times nT}$ are two block-lower triangular matrices representing system responses.

It has been proved in Ref.\cite{chen2023robust,ANDERSON2019364} that for any $\mathbf{\mathcal{T}}_{\boldsymbol{g}},\mathbf{\mathcal{T}}_{\textbf{u}}$ satisfying
\begin{equation}
    [\mathbb{I}-Z\mathbf{\mathcal{A}}\ \ -Z\mathbf{\mathcal{B}}]\left[
  \begin{gathered}
 \mathbf{\mathcal{T}}_{\boldsymbol{g}}\\
  \mathbf{\mathcal{T}}_\textbf{u}
  \end{gathered}
  \right]=\mathbb{I}, \label{the_desired_response_controller}
\end{equation}
the controller $\mathbf{\mathcal{T}}_\textbf{u}\mathbf{\mathcal{T}}_{\boldsymbol{g}}^{-1}\in\mathcal{L}^{mT\times nT}$ achieves the desired response. In Eq.\eqref{the_desired_response_controller}, $\mathbf{\mathcal{A}}$ is formed by diagonally concatenating $T$ instances of $\boldsymbol{A}$ along with a $p \times p$ zero matrix, expressed as $\mathbf{\mathcal{A}}=\text{blkdiag}(\boldsymbol{A},\cdots,\boldsymbol{A},\boldsymbol{0}_{p \times p})$. Similarly, $\mathbf{\mathcal{B}}$ is constructed through diagonal concatenation of $T$ instances of $\boldsymbol{B}$ and a $p \times q$ zero matrix, expressed as $\mathbf{\mathcal{B}}=\text{blkdiag}(\boldsymbol{B},\cdots,\boldsymbol{B},\boldsymbol{0}_{p \times q})$. $Z$ is the block-downshift operator, i.e., a matrix with the identity matrix on the first block subdiagonal and zeros elsewhere.

Based on the definition above, we further examine the effect of inaccurate Koopman linear representations $\bar{\boldsymbol{A}},\bar{\boldsymbol{B}}$ and $\bar{\boldsymbol{g}}$ on the controlled system dynamics and provide an estimation of the upper error bound given as Eq. \eqref{eq_identified_response}-Eq. \eqref{eq_and}. 

For the identified model $\Bar{\mathbf{\mathcal{A}}}=\text{blkdiag}(\Bar{\boldsymbol{A}},\cdots,\Bar{\boldsymbol{A}},\boldsymbol{0}_{p \times p})$ and $\Bar{\mathbf{\mathcal{B}}}=\text{blkdiag}(\Bar{\boldsymbol{B}},\cdots,\Bar{\boldsymbol{B}},\boldsymbol{0}_{p \times q})$, the block-lower triangular matrices $\{\Bar{\mathbf{\mathcal{T}}}_{\boldsymbol{g}},\Bar{\mathbf{\mathcal{T}}}_\textbf{u}\}$ satisfying:
\begin{align}
    [\mathbb{I}-Z\Bar{\mathbf{\mathcal{A}}}\ \ -Z\Bar{\mathbf{\mathcal{B}}}]\left[
  \begin{gathered}
 \Bar{\mathbf{\mathcal{T}}}_{\boldsymbol{g}}\\
  \Bar{\mathbf{\mathcal{T}}}_\textbf{u}
  \end{gathered}
  \right]=\mathbb{I} \label{eq_identified_response}
\end{align}
By rewriting Eq.\eqref{eq_identified_response}, we can obtain
\begin{align}
    [\mathbb{I}-Z\mathbf{\mathcal{A}}\ \ -Z\mathbf{\mathcal{B}}]
  \Bar{\mathbf{\mathcal{T}}}
  =\mathbb{I}-Z\boldsymbol{\Delta}\Bar{\mathbf{\mathcal{T}}} 
\end{align}
where $\boldsymbol{\Delta}=Z[\boldsymbol{\Delta}_{\mathcal{A}}\ \boldsymbol{\Delta}_{\mathcal{B}}], \Bar{\mathbf{\mathcal{T}}}^{\top}=[\Bar{\mathbf{\mathcal{T}}}_{\boldsymbol{g}}^{\top}\ \Bar{\mathbf{\mathcal{T}}}_{\textbf{u}}^{\top}]$ and $\boldsymbol{\Delta}_{\mathcal{A}},\boldsymbol{\Delta}_{\mathcal{B}}$ are block diagonal matrixs satisfying $\boldsymbol{\Delta}_{\mathcal{A}}=\Bar{\mathbf{\mathcal{A}}}-\mathbf{\mathcal{A}}, \boldsymbol{\Delta}_{\mathcal{B}}=\Bar{\mathbf{\mathcal{B}}}-\mathbf{\mathcal{B}}$. The response of the system $(\mathbf{\mathcal{A}},\mathbf{\mathcal{B}})$ with the controller $\Bar{\mathbf{\mathcal{T}}}_{\textbf{u}}\Bar{\mathbf{\mathcal{T}}}_{\boldsymbol{g}}^{-1}$ is given by
\begin{equation}
  \left[
  \begin{gathered}
  \boldsymbol{\mathcal{G}}^{\top}(\boldsymbol{x})\\
  \textbf{u}_{*}\\
  \end{gathered}
  \right]=(\Bar{\mathbf{\mathcal{T}}}+\Bar{\mathbf{\mathcal{T}}}\boldsymbol{\Delta}(\mathbb{I}-\Bar{\mathbf{\mathcal{T}}}\boldsymbol{\Delta})^{-1}\Bar{\mathbf{\mathcal{T}}})\boldsymbol{\gamma}
\end{equation}

We decompose $\mathbf{\mathcal{T}},\boldsymbol{\Delta}$, and $\boldsymbol{\gamma}$ as follows to separate the effects of the known initial condition $x_0$ from the unknown future disturbances $\gamma_{0:T-1}$:
\begin{align}
    \mathbf{\mathcal{T}}=\left[
  \begin{gathered}
 \Bar{\mathbf{\mathcal{T}}}_{\boldsymbol{g}}\\
  \Bar{\mathbf{\mathcal{T}}}_\textbf{u}
  \end{gathered}
  \right]=\left[\Bar{\mathbf{\mathcal{T}}}^0\ |\ \Bar{\mathbf{\mathcal{T}}}^{\Tilde{\boldsymbol{\gamma}}}\right]
\end{align}
\begin{align}
    \boldsymbol{\gamma}=\left[
  \begin{gathered}
 x_0\\ \Tilde{\boldsymbol{\gamma}}
  \end{gathered}
  \right], \boldsymbol{\Delta}= \left[
  \begin{gathered}
 \boldsymbol{\Delta}^0\\
  \boldsymbol{\Delta}^{\Tilde{\boldsymbol{\gamma}}}
  \end{gathered}
  \right]
\end{align}
where $\Bar{\mathbf{\mathcal{T}}}^0$ is the first block column of $\mathbf{\mathcal{T}}$, $\boldsymbol{\Delta}^0$ is the first block row of $\boldsymbol{\Delta}$.
Then we have
\begin{align}
   \left[
  \begin{gathered}
 \Bar{\boldsymbol{\mathcal{G}}}(\boldsymbol{x})\\ \Bar{\mathbf{u}}_{*}
  \end{gathered}
  \right] &-\left[
  \begin{gathered}
 \boldsymbol{\mathcal{G}}(\boldsymbol{x})\\ \mathbf{u}_{*}
  \end{gathered}
  \right] \notag \\ &=(\Bar{\mathbf{\mathcal{T}}}-\mathbf{\mathcal{T}})\boldsymbol{\gamma}+\Bar{\mathbf{\mathcal{T}}}^{\Tilde{\boldsymbol{\gamma}}} \boldsymbol{\Delta}^{\Tilde{\boldsymbol{\gamma}}}(\mathbb{I}-\Bar{\mathbf{\mathcal{T}}}^{\Tilde{\boldsymbol{\gamma}}}\boldsymbol{\Delta}^{\Tilde{\boldsymbol{\gamma}}})^{-1}\Bar{\mathbf{\mathcal{T}}}\boldsymbol{\gamma} \label{open_loop}
\end{align}
where
\begin{align}
    [\mathbb{I}-Z\mathbf{\mathcal{A}}\ \ \ -Z\mathbf{\mathcal{B}}](\Bar{\mathbf{\mathcal{T}}}-\mathbf{\mathcal{T}})=\boldsymbol{\Delta}\Bar{\mathbf{\mathcal{T}}}=\Bar{\mathbf{\mathcal{T}}}^{\Tilde{\boldsymbol{\gamma}}}\boldsymbol{\Delta}^{\Tilde{\boldsymbol{\gamma}}}
\end{align}
and
\begin{align}
    (\mathbb{I}-\Bar{\mathbf{\mathcal{T}}}^{\Tilde{\boldsymbol{\gamma}}}\boldsymbol{\Delta}^{\Tilde{\boldsymbol{\gamma}}})^{-1}=\sum_{k=0}^{k}(\Bar{\mathbf{\mathcal{T}}}^{\Tilde{\boldsymbol{\gamma}}}\boldsymbol{\Delta}^{\Tilde{\boldsymbol{\gamma}}})^T \label{eq_and}
\end{align}
where
\begin{align}
\left\|\boldsymbol{\Delta}^{\Tilde{\boldsymbol{\gamma}}}\right\|\le\left\|\boldsymbol{\Delta}\right\|\le\left\|\boldsymbol{\Delta}_{\mathcal{A}}\ \boldsymbol{\Delta}_{\mathcal{B}}\right\|\le \varepsilon_{\boldsymbol{A}}+\varepsilon_{\boldsymbol{B}}
\end{align}

Therefore, it can be concluded that when $\varepsilon_{\boldsymbol{A}}$ and $\varepsilon_{\boldsymbol{B}}$ converge to 0, $\left\|[
 \Bar{\boldsymbol{\mathcal{G}}}^{\top}(\boldsymbol{x})\ \Bar{\mathbf{u}}_{*}^{\top}
]^{\top}-[
 \boldsymbol{\mathcal{G}}^{\top}(\boldsymbol{x})\ \mathbf{u}_{*}^{\top}
]^{\top}\right\|$ converges to 0. In other words, the error of the open-loop dynamics is limited by the representation errors of Koopman eigenpairs.

\subsection{Safety margin tuning}
\label{Safety_margin_tuning}

Even though there is only a small error between the optimal control strategy obtained from Eq. \eqref{optimal_control_problem} and the actual optimal amount of load shedding, but after rounded to the nearest feasible value, they may be rounded to different feasibles values, which causes a discrepancy $d$ of the quantization interval. Consequently, it's important to revise the optimal control strategy in Eq.\eqref{optimal_control_problem} to ensure that the system frequency trajectory remains within the prescribed hard limits. In this section, we further design the constraints in Eq. \eqref{optimal_control_problem} to prevent the rounded optimal control strategy from causing the system frequency to exceed the prescribed hard limits.

\vspace{0.2cm}
\noindent \textbf{Proposition 1.} By replacing the frequency limits in Eq. \eqref{optimal_control_problem} with Eq. \eqref{eq_zeta}, it ensures that the optimal control strategy obtained from KLS, when rounded to the nearest value, does not violate the prescribed hard limits on the system frequency.
\begin{align}
\bar{\omega}_{t}\ge{\omega}_{\text{min}}+\zeta \label{eq_zeta}
\end{align}
where $\zeta$ satisfies
\begin{align}
  \zeta & \ge \left\| \boldsymbol{C}\sum\limits_{k=0}^{t-1}{{{\boldsymbol{A}}^{(t-1)-k}}}\boldsymbol{B} \right\|\frac{d}{2} \notag \\
 & +\underset{\alpha \in \mathfrak{A},\beta\in \mathfrak{B},\boldsymbol{u}\in \mathcal{U}}{\mathop{\max }}\,\left\| \bar{\omega }_{t}^{\alpha ,\beta }(\boldsymbol{u})-\omega_{t}^{\alpha ,\beta }(\boldsymbol{u}) \right\|  \label{zeta_satisfies}
\end{align}
where $\boldsymbol{C}=[C_1, C_2,..., C_p]\in\mathbb{R}^{p} $ is an observation matrix with the first element $C_1$ equal to 1 and the remaining elements equal to 0.

\vspace{0.1cm}
\textit{\textbf{Remark 6}}: The values of $\mathbf{\omega }_{t}^{\alpha ,\beta }(\cdot )$ can be obtained through power system simulation experiments, while the values of $\mathbf{\bar{\omega }}_{t}^{\alpha ,\beta }(\cdot )$ can be calculated using the linear prediction system Eq.\eqref{eq_linear_dynamic_system_equation}. Since $\alpha ,\beta $ and $\boldsymbol{u}$ are in the training set for Eq.\eqref{eq_linear_dynamic_system_equation}, the last term on the right hand side of Eq.\eqref{zeta_satisfies} is the largest prediction error of Eq.\eqref{eq_linear_dynamic_system_equation} on the frequency trajectories in the training set. Hence, the last term on the right hand side of Eq.\eqref{zeta_satisfies} is acquired upon the completion of the training process for Eq.\eqref{eq_linear_dynamic_system_equation}. 

\vspace{0.2cm}
\noindent \textit{Proof}. The constraints in the optimal control problem Eq. \eqref{optimal_control_problem} guarantee that the minimum value of $\bar{\omega }_{t}^{\alpha ,\beta }(\boldsymbol{\bar{u}}_{*}^{\alpha ,\beta })$ is no less than ${{\omega }_{\min }}+\varsigma $, and the steady-state value is no less than ${{\omega }_{\infty \min }}+\varsigma $. Thus, it is crucial to find an upper bound on the difference between $\bar{\omega }_{t}^{\alpha ,\beta }(\boldsymbol{\bar{u}}_{*}^{\alpha ,\beta })$ and $\omega_{t}^{\alpha ,\beta }({{Q}_{d}}(\boldsymbol{\bar{u}}_{*}^{\alpha ,\beta }))$ in order to determine the value of $\zeta$. The estimation of this upper bound is given as follows.
\begin{align}
  & \left\| \bar{\omega }_{t}^{\alpha ,\beta }(\boldsymbol{\bar{u}}_{*}^{\alpha ,\beta })-\omega_{t}^{\alpha ,\beta }({{Q}_{d}}(\boldsymbol{\bar{u}}_{*}^{\alpha ,\beta })) \right\| \notag \\
 & \le \left\| \bar{\omega }_{t}^{\alpha ,\beta }(\boldsymbol{\bar{u}}_{*}^{\alpha ,\beta })-\bar{\omega }_{t}^{\alpha ,\beta }({{Q}_{d}}(\boldsymbol{\bar{u}}_{*}^{\alpha ,\beta })) \right\| \notag  \\
 & +\left\| \bar{\omega }_{t}^{\alpha ,\beta }({{Q}_{d}}(\boldsymbol{\bar{u}}_{*}^{\alpha ,\beta }))-\omega_{t}^{\alpha ,\beta }({{Q}_{d}}(\boldsymbol{\bar{u}}_{*}^{\alpha ,\beta })) \right\| \notag  \\
 & \le \left\| \boldsymbol{C}\sum\limits_{k=0}^{t-1}{{{\boldsymbol{A}}^{(t-1)-k}}}\boldsymbol{B} \right\|\frac{d}{2} \notag  \\
 & +\left\| \bar{\omega }_{t}^{\alpha ,\beta }({{Q}_{d}}(\boldsymbol{\bar{u}}_{*}^{\alpha ,\beta }))-\omega_{t}^{\alpha ,\beta }({{Q}_{d}}(\boldsymbol{\bar{u}}_{*}^{\alpha ,\beta })) \right\| \\
 & \le \left\| \boldsymbol{C}\sum\limits_{k=0}^{t-1}{{{\boldsymbol{A}}^{(t-1)-k}}}\boldsymbol{B} \right\|\frac{d}{2} \notag  \\
 & + \underset{\alpha \in \mathfrak{A},\beta\in \mathfrak{B},\boldsymbol{u}\in \mathcal{U}}{\mathop{\max }}\,\left\| \bar{\omega }_{t}^{\alpha ,\beta }(\boldsymbol{u})-\omega_{t}^{\alpha ,\beta }(\boldsymbol{u}) \right\|  
 \label{zeta_proof}
\end{align}
where $\boldsymbol{\bar{u}}_{*}^{\alpha,\beta}$ represents the optimal load shedding solution obtained by solving Eq. \eqref{optimal_control_problem}, while ${{Q}_{d}}(\boldsymbol{\bar{u}}_{*}^{\alpha, \beta})$ denotes the actual load shedding amounts at each load node $i$, $\bar{\omega }_{t}^{\alpha, \beta}(\boldsymbol{\bar{u}}_{*}^{\alpha ,\beta })$ corresponds to the predicted frequency of the linear prediction system at time $t$ when the load shedding amount is $\boldsymbol{\bar{u}}_{*}^{\alpha ,\beta }$. $\omega_{t}^{\alpha ,\beta }({{Q}_{d}}(\boldsymbol{\bar{u}}_{*}^{\alpha ,\beta }))$, and $\bar{\omega }_{t}^{\alpha ,\beta }({{Q}_{d}}(\boldsymbol{\bar{u}}_{*}^{\alpha ,\beta }))$ respectively represent the actual and predicted system frequency at time $t$ when the load shedding amount is ${{Q}_{d}}(\bar{u}_{*}^{\alpha ,\beta })$. The values of $\omega_{t}^{\alpha ,\beta }(\cdot)$ can be obtained through power system simulation, while the values of $\bar{\omega}_{t}^{\alpha ,\beta }(\cdot)$ can be calculated using the linear prediction system Eq.\eqref{eq_linear_dynamic_system_equation}.

The proof for the second inequality in Eq.\eqref{zeta_proof} is as follows.

\begin{align}
    \bar{\boldsymbol{\omega}}_{t}(\bar{\mathbf{u}})- 
    \bar{\boldsymbol{\omega}}_{t}(Q(\bar{\mathbf{u}}))&=\boldsymbol{C}\bar{\boldsymbol{A}}^{t}\boldsymbol{g}(\boldsymbol{x}_0)+\boldsymbol{C}\sum_{k=0}^{t-1}\bar{\boldsymbol{A}}^{(t-1)-k}\bar{\boldsymbol{B}}\boldsymbol{u}_{t} \notag  \\
    &-\boldsymbol{C}\bar{\boldsymbol{A}}^{t}\boldsymbol{g}(\boldsymbol{x}_0)-\boldsymbol{C}\sum_{k=0}^{t-1}\bar{\boldsymbol{A}}^{(t-1)-k}\bar{\boldsymbol{B}}Q(\boldsymbol{u}_{t}) \notag \\
    &=\boldsymbol{C}\sum_{k=0}^{t-1}\bar{\boldsymbol{A}}^{(t-1)-k}\bar{\boldsymbol{B}}(\boldsymbol{u}_{t}-Q(\boldsymbol{u}_{t})) \notag \\
    &\le\left\|\boldsymbol{C}\sum_{k=0}^{t-1}\bar{\boldsymbol{A}}^{(t-1)-k}\bar{\boldsymbol{B}}\right\|\frac{d}{2} \label{The_proof_for_the_second_inequality}
\end{align}

The safety margin $\zeta$ is expected to ensure that the KLS guarantees the system frequency to remain within the safe range under anticipated operating conditions $\alpha\in\mathfrak{A}$ and faults $\beta\in\mathfrak{B}$. 

Therefore, when the inequality Eq. \eqref{eq_zeta} holds, it ensures that the frequency trajectory of the actual system is within the safety range, under the load shedding amount ${{Q}_{d}}(\boldsymbol{\bar{u}}_{*}^{\alpha ,\beta })$.
$\hfill\blacksquare$


\vspace{0.1cm}
\textit{\textbf{Remark 7}}: 
Manual adjustment of $\zeta$ is possible. This entails the following steps. First, simulating the system frequency by rounding the optimal control strategy to the nearest feasible value for each operating condition and fault in the training set. Then, instances where the system frequency fails to meet the hard limits are selected. $\zeta$ is then increased and the system frequency is simulated under the new $\zeta$. This step is repeated until the system frequency complies with the limits. On the contrast, 
our approach avoids the extensive simulation required to find suitable values of $\zeta$. Instead, it relies on $\boldsymbol{A}, \boldsymbol{B}$ in Eq.\eqref{eq_linear_dynamic_system_equation}, and the prediction error already computed during the training of the encoder, thus enhancing the efficiency of $\zeta$ design.

Here, we further discuss when the equality holds in the inequality \eqref{zeta_proof}. 

In the deviation of the upper bound for $\left\| \bar{\omega }_{t}^{\alpha ,\beta }(\boldsymbol{\bar{u}}_{*}^{\alpha ,\beta })-\omega_{t}^{\alpha ,\beta }({{Q}_{d}}(\boldsymbol{\bar{u}}_{*}^{\alpha ,\beta })) \right\|$, the first inequality in \eqref{zeta_proof} and the last inequality in \eqref{The_proof_for_the_second_inequality} are based on the sub-multiplicative inequality and the triangle inequality of the Frobenius norm, respectively. For any two arrays $\boldsymbol{\mathcal{M}}$ and $\boldsymbol{\mathcal{N}}$, equality for the triangle inequality holds when the two arrays are linearly dependent, while equality for the sub-multiplicative inequality holds if and only if each row of $\boldsymbol{\mathcal{M}}$ and each column of $\boldsymbol{\mathcal{N}}$ are linearly dependent. 

$\left\| \bar{\omega }_{t}^{\alpha ,\beta }(\boldsymbol{\bar{u}}_{*}^{\alpha ,\beta })-\omega_{t}^{\alpha ,\beta }({{Q}_{d}}(\boldsymbol{\bar{u}}_{*}^{\alpha ,\beta })) \right\|$ is often strictly lower than the upper bound derived in Eq.\eqref{zeta_proof}. The reason is the equality conditions of the triangle inequality and the sub-multiplicative inequality in \eqref{zeta_proof} and \eqref{The_proof_for_the_second_inequality} are hard to satisfy. The gap between $\left\| \bar{\omega }_{t}^{\alpha ,\beta }(\boldsymbol{\bar{u}}_{*}^{\alpha ,\beta })-\omega_{t}^{\alpha ,\beta }({{Q}_{d}}(\boldsymbol{\bar{u}}_{*}^{\alpha ,\beta })) \right\|$ and its upper bound will be further illustrated in the simulation results in Section \ref{Safety Margin}.


\section{Case Study} \label{case_study}
In this section, the prediction capability and the control effect of KLS is illustrated by a case study on the CloudPSS platform \cite{SONG20201611}, \cite{CloudPSS}. 

\subsection{Test System and Datasets} \label{Test_System_Description}
To validate the effectiveness of our proposed method, we conducted simulation experiments on the CIGRE-LF test system, which is adapted from a real provincial power grid in China. The CIGRE-LF consists of 102 500-kV buses, and possesses a load level of 2600 MW, with installed capacities of 2400 MW and 5400 MW for renewable and conventional energy sources, respectively. The full electromagnetic transient (EMT) model of the test system is built on the CloudPSS platform \cite{8582334}.

Based on the synchronous generators' model, system inertia is an important parameter that affects frequency safety. The influence of operating conditions on frequency dynamics can be attributed, in part, to the variations in system inertia caused by changes in operating conditions. Therefore, in this section, we assume a certain level of randomness in system inertia to capture unanticipated operating conditions. 

The training and testing sets are generated as follows. In CEPRI-LF, synchronous generators in a region are represented as an equivalent single unit. Variations in system operating conditions may result in changes in the unit commitment, potentially leading to variations in the inertia of the equivalent unit. To simulate the variation in system operation, we assume that the parameter values of inertia for each equivalent unit follow a uniform distribution, with the mean being the original manufacturer data and a deviation of 30\% of the mean value to account for parameter uncertainties. We assume that ${u}_{i}$ at bus $i$ is a uniformly distributed random number between 0 and 1. The fault set of the system is generated by traversing N-1, N-2, and N-3 faults. The training set consists of 600 frequency trajectories, while the testing set consists of 300 frequency trajectories. Each frequency trajectory has a length of 1 min. When generating a frequency trajectory, inertia for each equivalent generator, ${u}_{i}$, and faults are randomly generated according to their respective distributions. Since ${u}_{i}$ and the inertia for each equivalent generator are continuous variables, the probability of having the same ${u}_{i}$ and inertia values for two frequency trajectories is small. Therefore, we assume that there is no intersection between the training and testing sets.

\subsection{Adaptability}
\begin{figure}[t] 
    \centering
    \setlength{\abovecaptionskip}{-0.05cm}   
    \setlength{\belowcaptionskip}{-2cm}   
    \includegraphics[width=8.5cm]{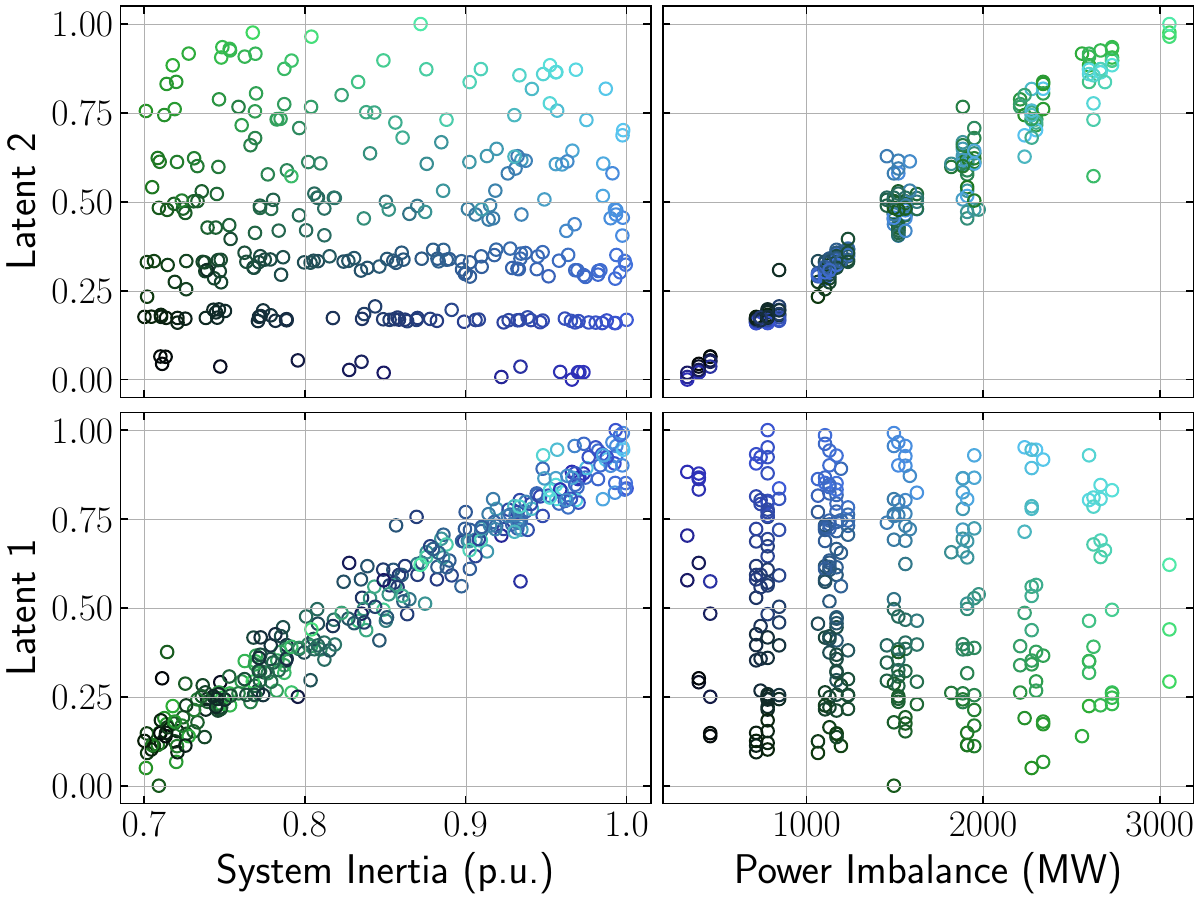}
    \caption{The correlation between the system inertia, power imbalance, and the outputs of the latent extractor.} \label{SCATTER_LATENTS_INRETIA_FAULT}
\end{figure}
Fig.\ref{SCATTER_LATENTS_INRETIA_FAULT} demonstrates that KLS is capable of extracting latent variables strongly correlated with the system inertia and power imbalance from the frequency trajectory within a 300 ms time window after a fault occurs. Here, the system inertia equals to 1 when the inertia for each equivalent unit equals to the corresponding original manufacturer data. Specifically, Latent1 and Latent2 exhibit a remarkably high correlation with the system inertia and power imbalance, respectively, providing evidence that Koopman linear representations are capable of capturing parameter variations in the original state space system model using measurements. This result validates the adaptability of KLS to diverse operating conditions and faults. Furthermore, as the results presented in Fig.\ref{SCATTER_LATENTS_INRETIA_FAULT} are based on the training set, it confirms that KLS is able to adapt to unanticipated operating conditions and faults.

\subsection{Prediction Capability}
\begin{figure}[t] 
    \centering
    \setlength{\abovecaptionskip}{-0.05cm}   
    \setlength{\belowcaptionskip}{-2cm}   
    \includegraphics[width=8.5cm]{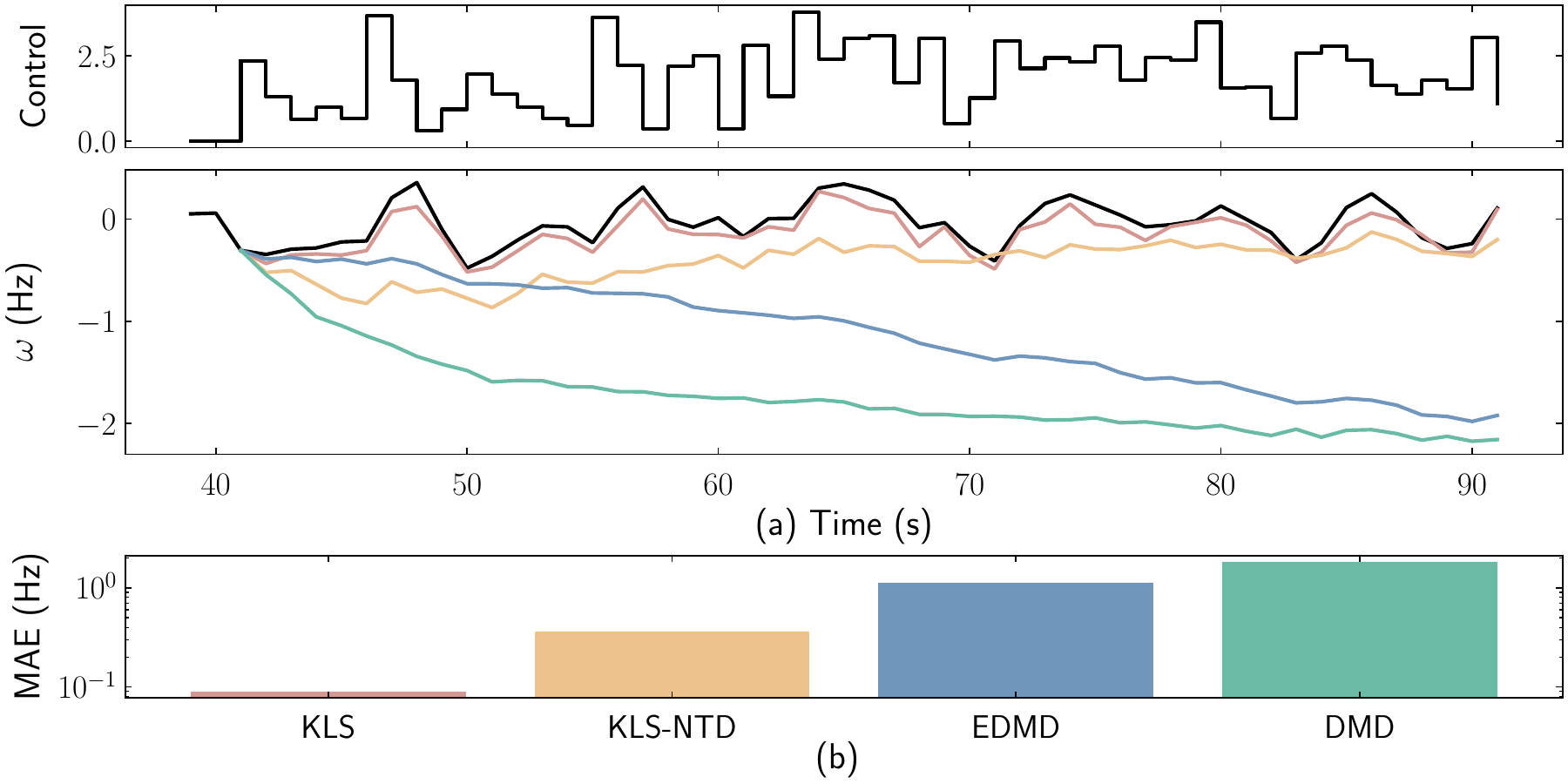}
    \caption{Comparison between true frequency trajectory(black) and predicted frequency trajectory(pink) during future 60s with proposed KLS(pink), KLS-NTD (orange), EDMD (blue) and DMD (green). (a) True and predicted frequency trajectories under random control inputs ; (b) MAE of prediction errors.} \label{PREDICTED_TRAJECTORY}
\end{figure}
The strong nonlinearity of dynamics makes the local linearization method hard to fit the model accurately in global horizon, and the piecewise linearization has difficulty in remaining a balance between accuracy and simplicity. Therefore, only Koopman operator related algorithms are compared. As the benchmark of the learning algorithm, the DMD and EDMD method is implemented with 100 radial basis functions (RBF) as observables. To illustrate the effectiveness of incorporating time delay embedding, KLS without time delay embedding (KLS-NTD, i.e., when $\tau=0$ in Eq.\eqref{time_delay}) is also tested as a benchmark. In subsequent discussions, we refer to KLS-NTD, DMD, and EDMD as the state-of-the-art algorithms (SOTAs) for brevity and clarity.

To illustrate the capability of KLS to learn the dynamics of system frequency from the online measurement, the frequency measurement of 1 min after a fault occurs is used to fit the linear model in Eq.\eqref{eq_linear_dynamic_system_equation} for the system with different inertia, control inputs and faults. 

Fig.\ref{PREDICTED_TRAJECTORY}(a) demonstrates the global linearization capability of the proposed methods. The pink line represents the simulated dynamic process, while the blue and green lines represent the predicted results obtained from EDMD and DMD, respectively. The orange line corresponds to the predicted results obtained from KLS-NTD.

Based on the frequency sequence observed within 300ms after the generator tripping at 40s, KLS demonstrates accurate prediction of the evolving frequency for the subsequent 60 seconds under different control inputs. This highlights the effectiveness of the latent extractor combined with time-delayed measurements in capturing the dynamics of the system frequency. In contrast, DMD and EDMD exhibit poorer performance, which can be attributed to their limited capability in incorporating time-delay information and harnessing the powerful non-linear representation offered by deep learning techniques.
\begin{figure}[t] 
    \centering
    \setlength{\abovecaptionskip}{-0.05cm}   
    \setlength{\belowcaptionskip}{-2cm}   
    \includegraphics[width=8.5cm]{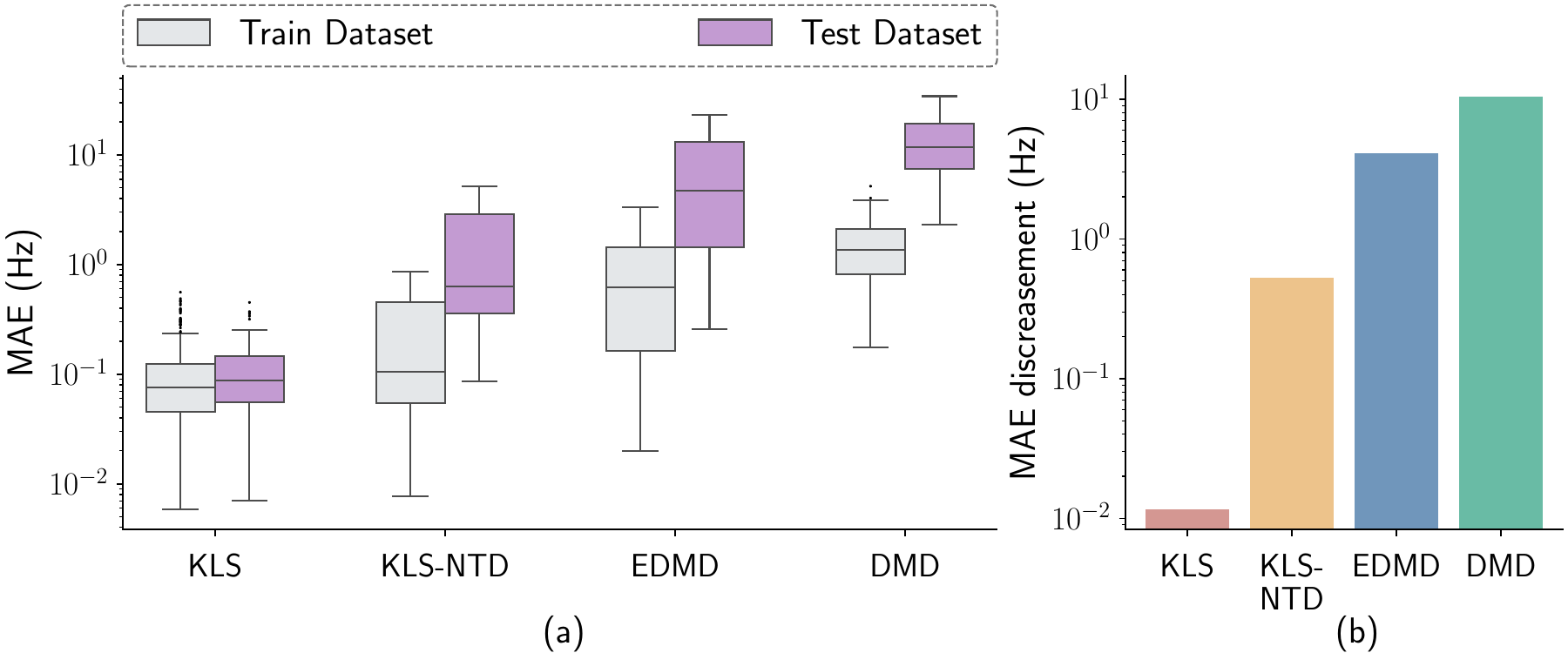}
    \caption{Boxplot of prediction errors of different Koopman operator models.} \label{BOXPLOT_SCATTER_COMPARISON_PRED_ERROR_2}
\end{figure}
The prediction accuracy of KLS and SOTAs on the training and testing sets is illustrated in Fig.\ref{BOXPLOT_SCATTER_COMPARISON_PRED_ERROR_2}. The mean average error (MAE) is employed as a measure of prediction accuracy. Due to the capability of KLS to track parameter variations in the original state space system model, the prediction accuracy is consistently high on both the training and testing sets.
\begin{figure}[t] 
    \centering
    \setlength{\abovecaptionskip}{-0.05cm}   
    \setlength{\belowcaptionskip}{-2cm}   
    \includegraphics[width=8.5cm]{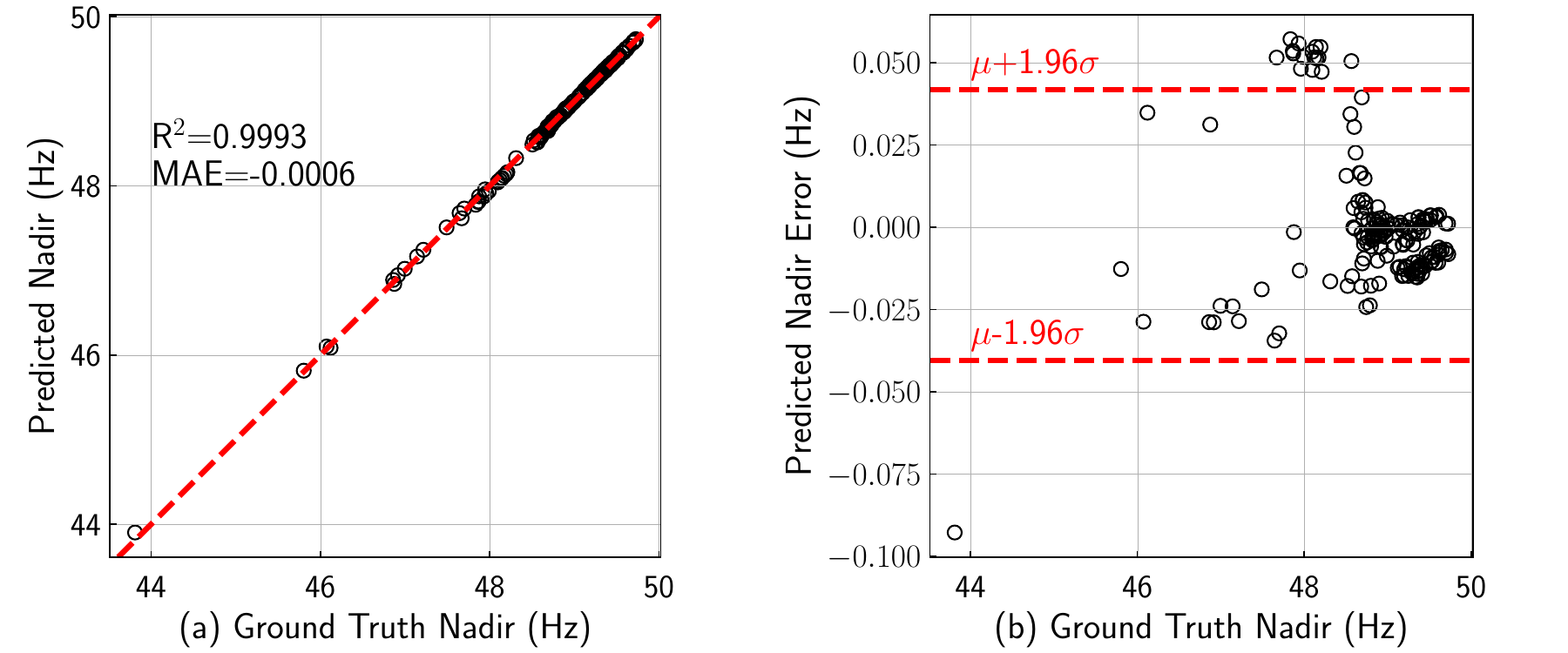}
    \caption{Prediction accuracy of KLS for the frequency nadirs.} \label{SCATTER_PRED_ERROR_NADIR}
\end{figure}

\begin{figure}[t] 
    \centering
    \setlength{\abovecaptionskip}{-0.05cm}   
    \setlength{\belowcaptionskip}{-2cm}   
    \includegraphics[width=8.5cm]{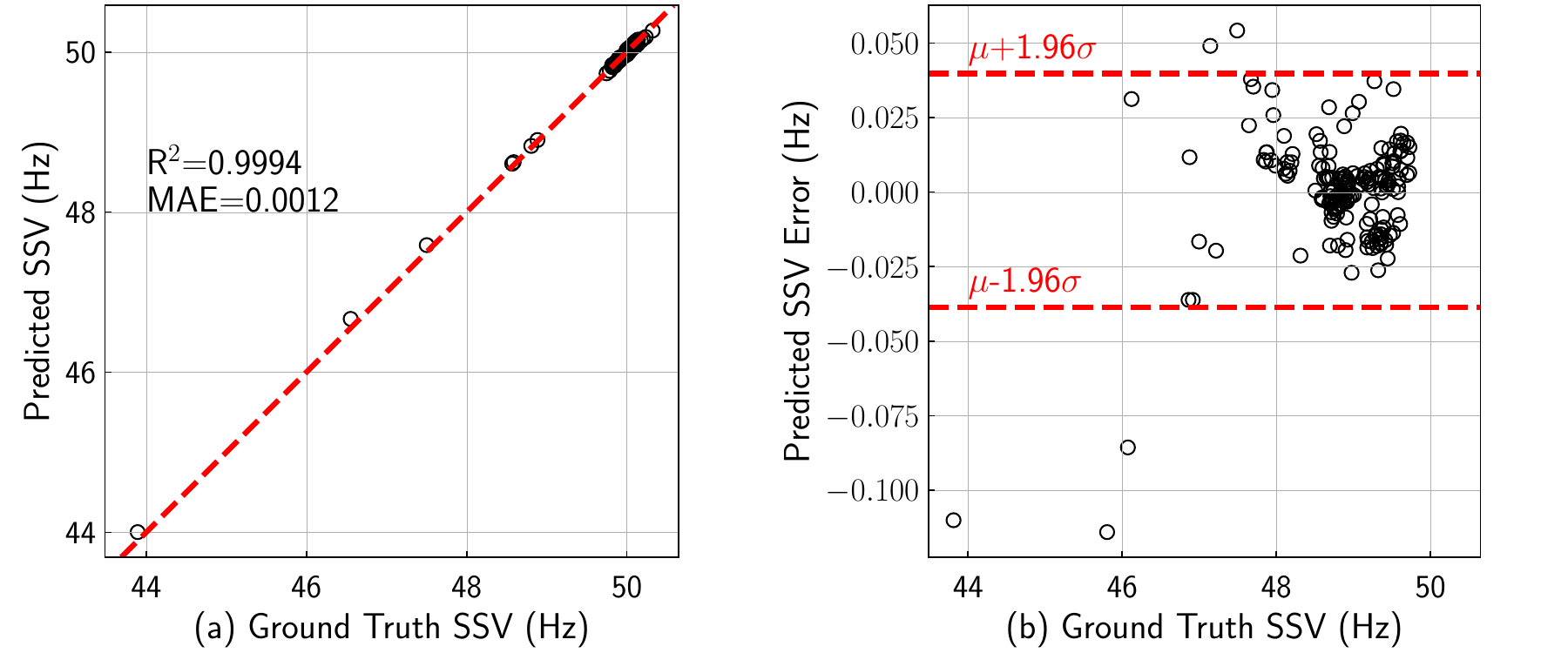}
    \caption{Prediction accuracy of KLS for the steady-state values.} \label{SCATTER_PRED_ERROR_SSV}
\end{figure}

For frequency control problems, the frequency nadir and steady-state value are two important indicators for assessing frequency safety. In order to demonstrate the prediction accuracy of KLS for the frequency nadir and steady-state value, we set all the inputs $\boldsymbol{u}$ in the test dataset to zero and re-simulated the true frequency trajectory. Subsequently, we computed the predicted frequency trajectory using KLS. The prediction accuracy for both indicators is shown in Fig. \ref{SCATTER_PRED_ERROR_NADIR} and Fig. \ref{SCATTER_PRED_ERROR_SSV}. It can be observed that the MAE of the predicted frequency trajectories for 95\% of the test dataset is within 0.1\%.

\subsection{Control Effect}
In this subsection, we focus on investigating the impact of prediction accuracy on control effectiveness. Hence, it is assumed that a continuous adjustment of load shedding can be achieved, and the safety margin $\zeta$ is set to zero. The effectiveness of KLS and SOTAs is evaluated by assessing the system's frequency safety after the implementation of these control strategies. Furthermore, to demonstrate the adaptability of KLS to unanticipated operating conditions and faults, all results in this subsection are computed using the test dataset.
The normalized safety metric is calculated as follows:
\begin{align}
    Safety=\alpha \left(\frac{Nadir^{*}-Nadir_{0}^{S}}{Nadir_{1}^{S}-Nadir_{0}^{S}}\right)+\beta \left(\frac{SSV^*-SSV_{0}^{S}}{SSV_{1}^{S}-SSV_{0}^{S}}\right) \label{normalized_safety_metric}
\end{align}

In Eq.\eqref{normalized_safety_metric}, when the frequency nadir is no less than $Nadir_{1}^{S}$ and the steady-state frequency is no less than $SSV_{1}^{S}$, $Safety$ is assigned a value of 1; when the system's frequency nadir $Nadir^{*}$ is less than $Nadir_{0}^{S}$ and the steady-state frequency $SSV^{*}$ is less than $SSV_{0}^{S}$, $Safety$ is assigned a value of 0. The weights for measuring the safety indicators of nadir and steady-state value are represented by $\alpha$ and $\beta$, respectively. In this paper, the values of $Nadir_1^{S}$ and $SSV_1^{S}$ are set to 49.0Hz and 49.5Hz, respectively, which are equal to $\omega_{\min}$ and $\omega_{\infty \min}$. $Nadir_0^{S}$ and $SSV_0^{S}$ are set to 48.5Hz and 49.0Hz, respectively. The values of $\alpha$ and $\beta$ are both set to 0.5.

\begin{figure}[t] 
    \centering
    \setlength{\abovecaptionskip}{-0.05cm}   
    \setlength{\belowcaptionskip}{-2cm}   
    \includegraphics[width=8.5cm]{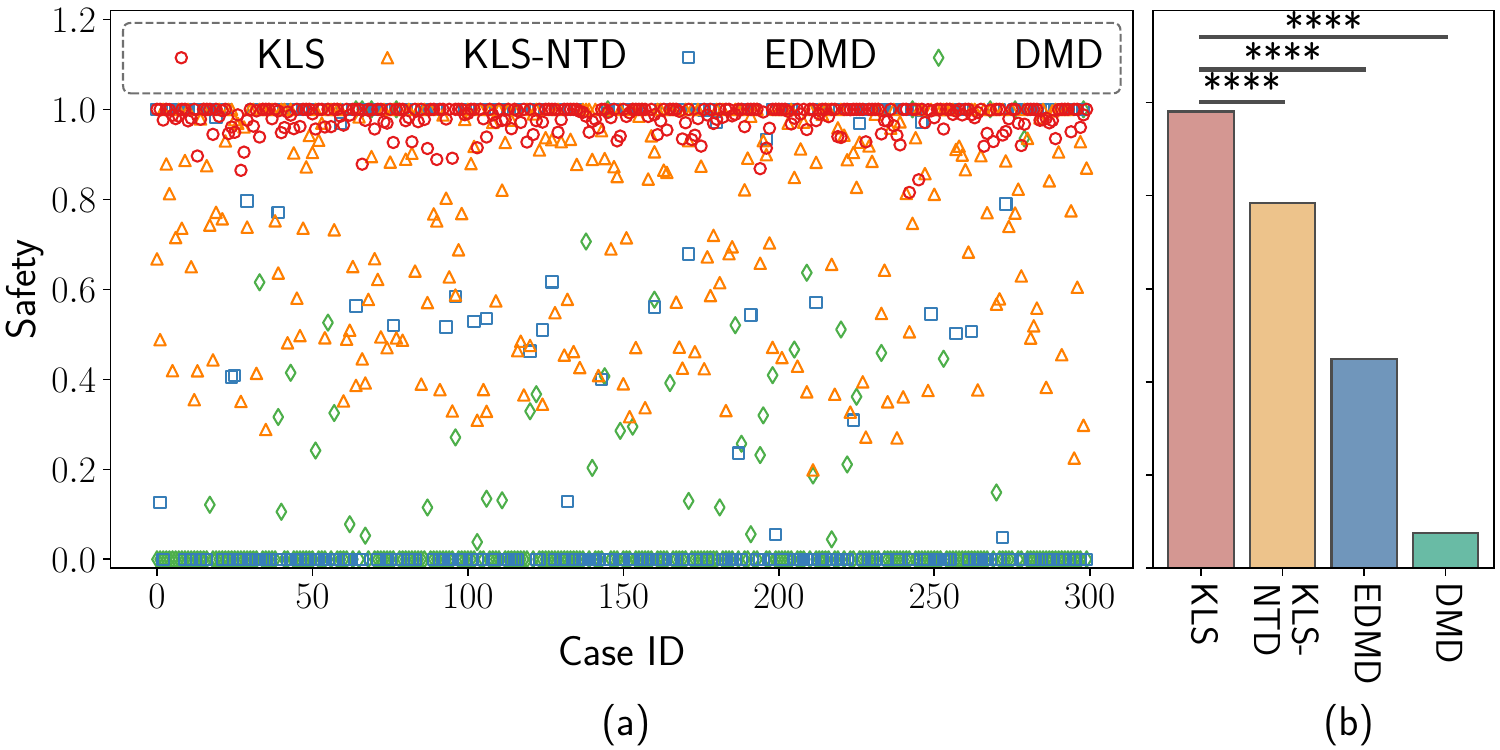}
    \caption{The average control safety metrics of KLS, KLS-NTD, EDMD, and DMD on the test dataset.} \label{COMPARISON_CONTROL_SAFETY}
\end{figure}

The average control safety metrics of KLS, KLS-NTD, EDMD, and DMD on the test dataset are depicted in Fig.\ref{COMPARISON_CONTROL_SAFETY}. T-tests are conducted to compare the effectiveness of different control strategies. The symbols **** represent T-test results with significance levels less than 0.0001. Among the 300 test data, the proportions of $Safety$ exceeding 0.9 for these methods are 97.5\%, 49.3\%, 41.6\%, and 4\% respectively. KLS, with its superior predictive accuracy compared to SOTAs, enhances the system's frequency safety. However, without incorporating a safety margin into the frequency constraints of the optimal control problem Eq. \eqref{optimal_control_problem}, the average control safety metric of KLS fails to reach a value of 1. In Fig.\ref{SCATTER_PRED_ERROR_NADIR} and Fig.\ref{SCATTER_PRED_ERROR_SSV}, approximately 2.5\% of the test data exhibit prediction errors exceeding +0.05Hz. Furthermore, KLS tends to overestimate the frequency nadir and steady-state frequency for these 2.5\% of the test data, resulting in relatively conservative control measures. As a result, around 2.5\% of the post-control frequency trajectories fall below the safety metric threshold of 0.9. Considering the feasible load shedding amounts and the discrepancy from the optimal control obtained through Eq.\eqref{optimal_control_problem}, further deterioration of KLS's $Safety$ would occur. This observation underscores the necessity of incorporating a safety margin in the frequency constraints of the optimal control problem.

\subsection{Safety Margin}
\label{Safety Margin}
This subsection analyzes the control effectiveness after the introduction of a safety margin in Subsection \ref{Safety_margin_tuning}. The evaluation of control measures is based on two indicators: the system's frequency safety and the control cost.
Increasing the amount of load shedding typically results in a higher system frequency. If the system frequency remains within the safe range (i.e., $Safety$ = 1), the greater the deviation of the nadir and the steady-state value of the system's frequency from the specified hard limits are, the higher the associated control cost becomes. Hence, a normalized economic metric $Control\ Cost$, ranging from 0 to 1, is introduced as an indicator to measure the control cost.
\begin{align}
    Control\ &Cost  \notag \\
    &=min\left(\frac{Nadir^*-Nadir_0^{E}}{Nadir_1^{E}-Nadir_0^{E}},\frac{SSV^*-SSV_0^{E}}{SSV_1^{E}-SSV_0^{E}}\right)
\end{align}
where $Nadir_1^{E}=49.0$Hz and $SSV_1^{E}=49.5$Hz indicate that when the nadir is equal to 49.0Hz and the steady-state value is equal to 49.5Hz, the $Control $ of load shedding measures is assigned a value of 1. Similarly, $Nadir_0^{E}=49.5Hz$ and $SSV_0^{E}=50.0$Hz indicate that when the nadir is no less than 49.5Hz and the steady-state value is no less than 50.0Hz, the $Control\ Cost$ of load shedding measures is assigned a value of 0.

\begin{figure}[t] 
    \centering
    \setlength{\abovecaptionskip}{-0.05cm}   
    \setlength{\belowcaptionskip}{-2cm}   
    \includegraphics[width=8.5cm]{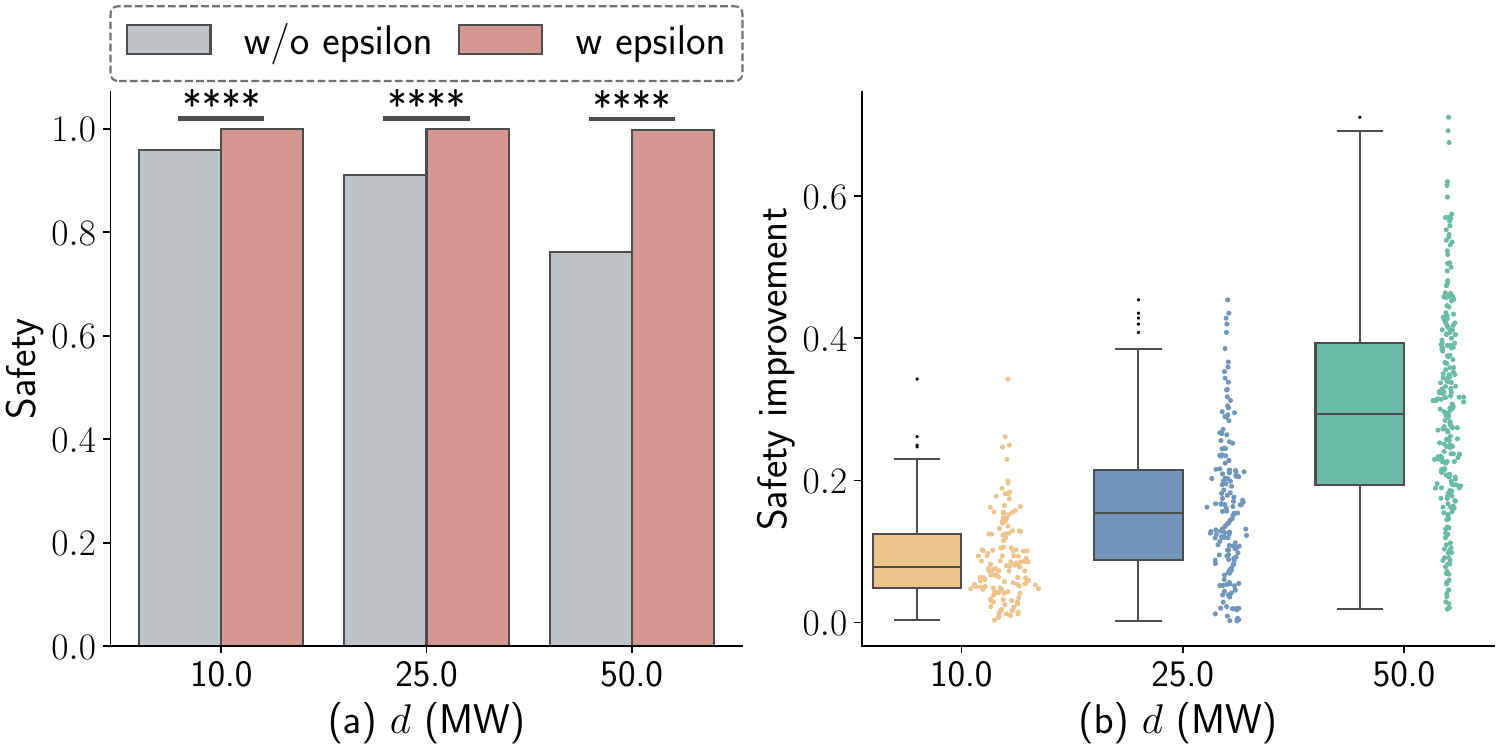}
    \caption{The improvement in $Safety$ index when incorporating the safety margin.} \label{BOXPLOT_SCATTER_CONTROL_W_WO_EPSILON}
\end{figure}

Fig. \ref{BOXPLOT_SCATTER_CONTROL_W_WO_EPSILON} presents the improvement in $Safety$ when incorporating the safety margin calculated using Eq.\eqref{optimal_control_problem} at different values of $d$ (where $d$ takes values of 10MW, 25MW, and 50MW).
It can be observed that the introduction of the safety margin enhances the system's frequency safety for different load levels associated with a feeder. Moreover, as $d$ increases, the improvement in $Safety$ becomes more significant.

\begin{figure}[t] 
    \centering
    \setlength{\abovecaptionskip}{-0.05cm}   
    \setlength{\belowcaptionskip}{-2cm}   
    \includegraphics[width=8.5cm]{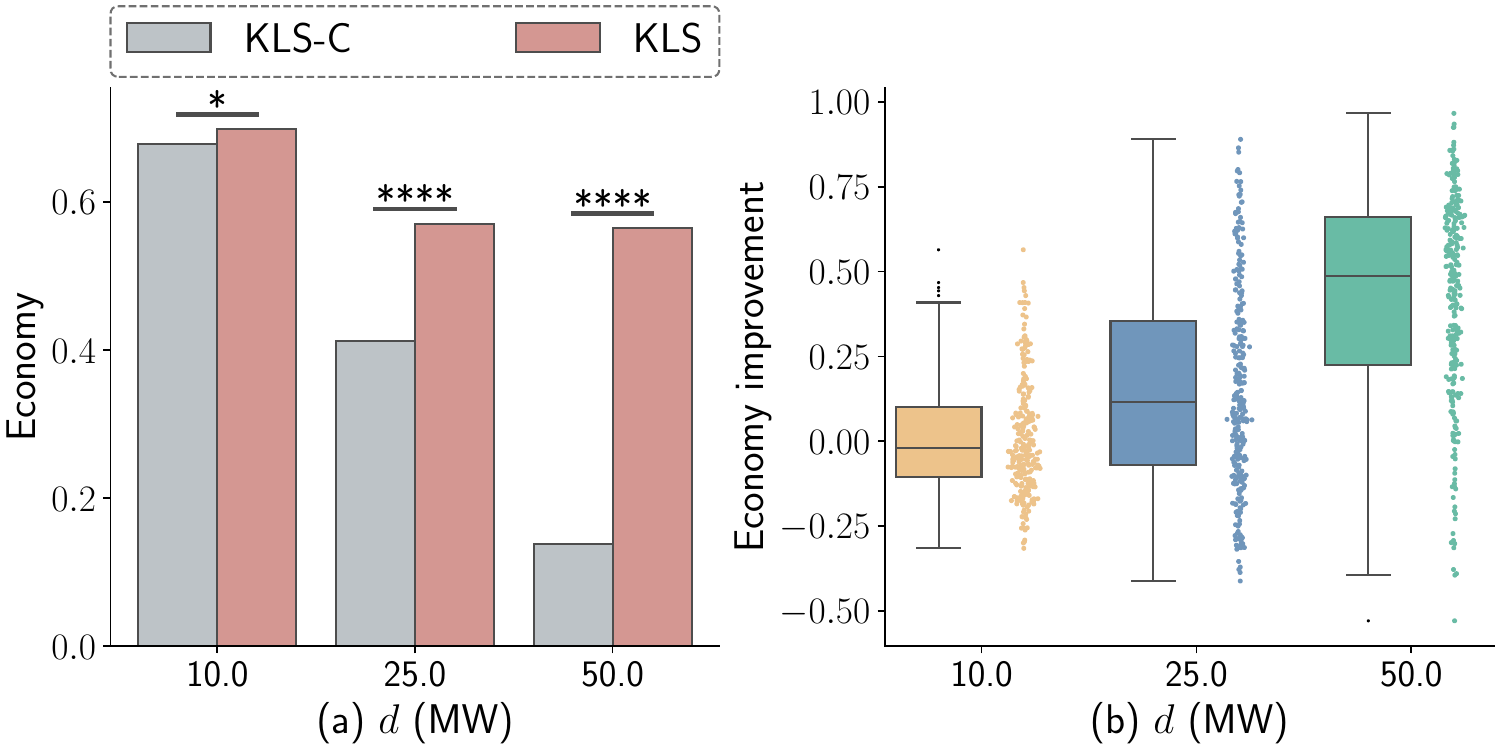}
    \caption{The $Control\ Cost$ index of KLS-C and KLS.} \label{BOXPLOT_SCATTER_CONTROL_EPSILON_CEIL}
\end{figure}

For the sake of clarity, we will refer to the control strategy 
in Eq. \eqref{actual_load_shedding_amount_larger} as the ceiled KLS (KLS-C). The $Control\ Cost$ of KLS-C and KLS is demonstrated in Fig. \ref{BOXPLOT_SCATTER_CONTROL_EPSILON_CEIL}. The symbols * represent T-test results with significance levels less than 0.1. It can be observed that by incorperating the load shedding amount given in Eq.\eqref{actual_load_shedding_amount} along with the safety margin, KLS achieves a reduced level of load shedding amount compared to the strategy in Eq.\eqref{actual_load_shedding_amount_larger}. This reduction is achieved while ensuring the safety of the system's frequency

\begin{figure}[t] 
    \centering
    \setlength{\abovecaptionskip}{-0.05cm}   
    \setlength{\belowcaptionskip}{-2cm}   
    \includegraphics[width=8.5cm]{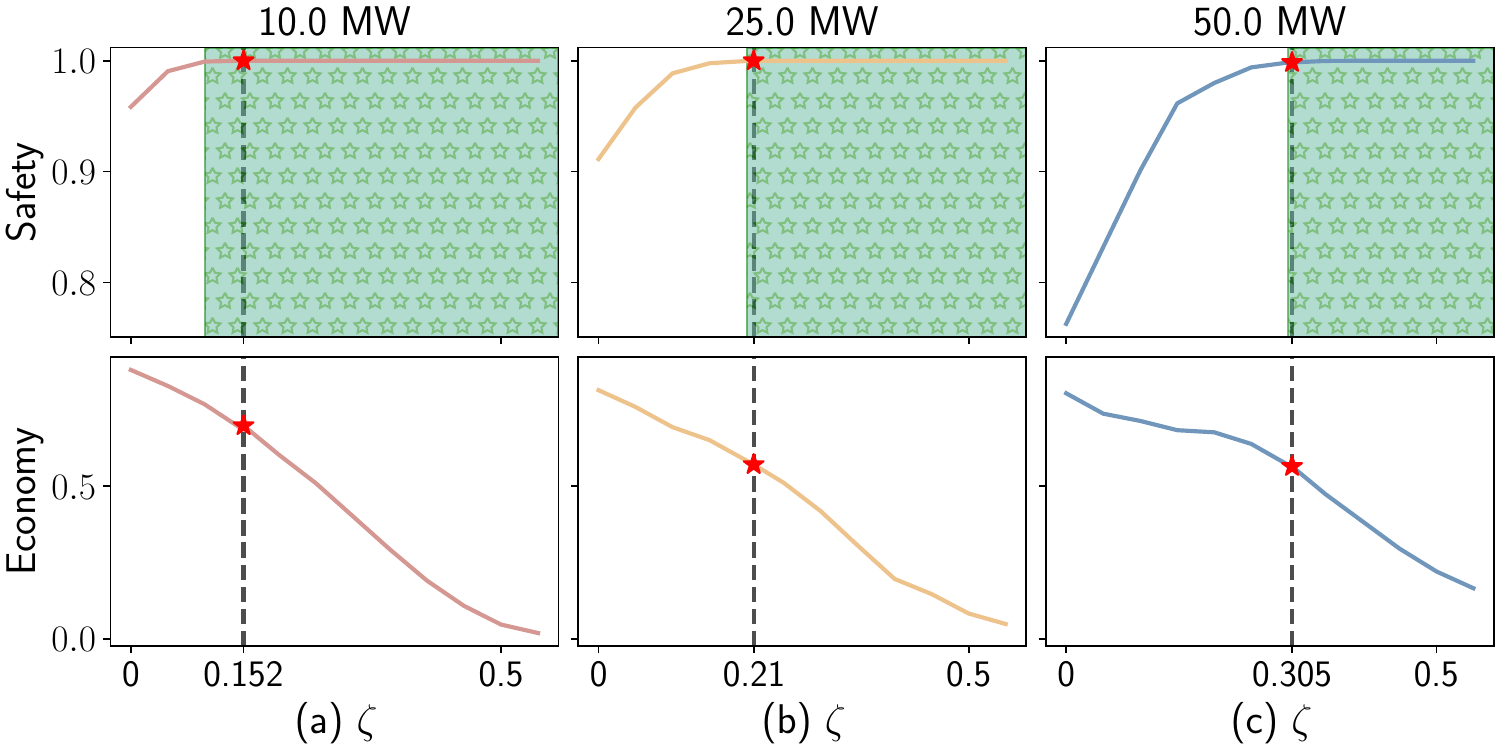}
    \caption{Comparison between the ideal safety margin and the safety margin calculated by Eq.\eqref{zeta_satisfies}.} \label{SCATTER_CURVE_CONTROL_EPSILON_IS_BEST}
\end{figure}

Fig. \ref{SCATTER_CURVE_CONTROL_EPSILON_IS_BEST} presents the safety margin calculated by Eq. \eqref{eq_zeta}. Firstly, under different values of $d$ (10MW, 25MW, and 50MW), we simulated the $Safety$ and $Control\ Cost$ index of KLS for various values of $\zeta$ ranging from 0 to 0.5. In Fig. \ref{SCATTER_CURVE_CONTROL_EPSILON_IS_BEST}, the green region represents the system satisfying $Safety=1$. The boundary between the green and white regions corresponds to an $\zeta$ that ensures system frequency safety, while reaching a minimum load shedding amount. Next, under different values of $d$ (10MW, 25MW, and 50MW), $\zeta$ was calculated using Eq. \eqref{zeta_satisfies}. The calculated values of minimal $\zeta$ obtained from Eq. \eqref{zeta_satisfies} are 0.152, 0.21, and 0.305, respectively. It can be observed that the $\zeta$ computed by the right hand side in \eqref{zeta_satisfies} are slightly larger than the ideal values obtained by iterating over different $\zeta$ values. This discrepancy arises from the difficulty in satisfying the inequality in Eq. \eqref{zeta_satisfies}. However, the $\zeta$ obtained by Eq. \eqref{zeta_satisfies} ensure the system's frequency safety.

The result of $\zeta$ provided by the right hand side of Eq. \eqref{zeta_satisfies} represents an upper bound determined by the prediction error of the Koopman linear system and the interval of discrtete load shedding amount. For a $\zeta$ smaller than the bound, the system frequency may fall out of the hard limits with the optimal strategy provided by KLS. Conversely, for a $\zeta$ larger than the bound, the optimal load shedding amount provided by KLS increases, leading to unnecessary economic losses. Hence, our safety margin tuning method provides a $\zeta$ that balance both safety and the control cost.

\subsection{Comparison with the traditional UFLS scheme}
\begin{figure}[t] 
    \centering
    \setlength{\abovecaptionskip}{-0.05cm}   
    \setlength{\belowcaptionskip}{-2cm}   
    \includegraphics[width=8.5cm]{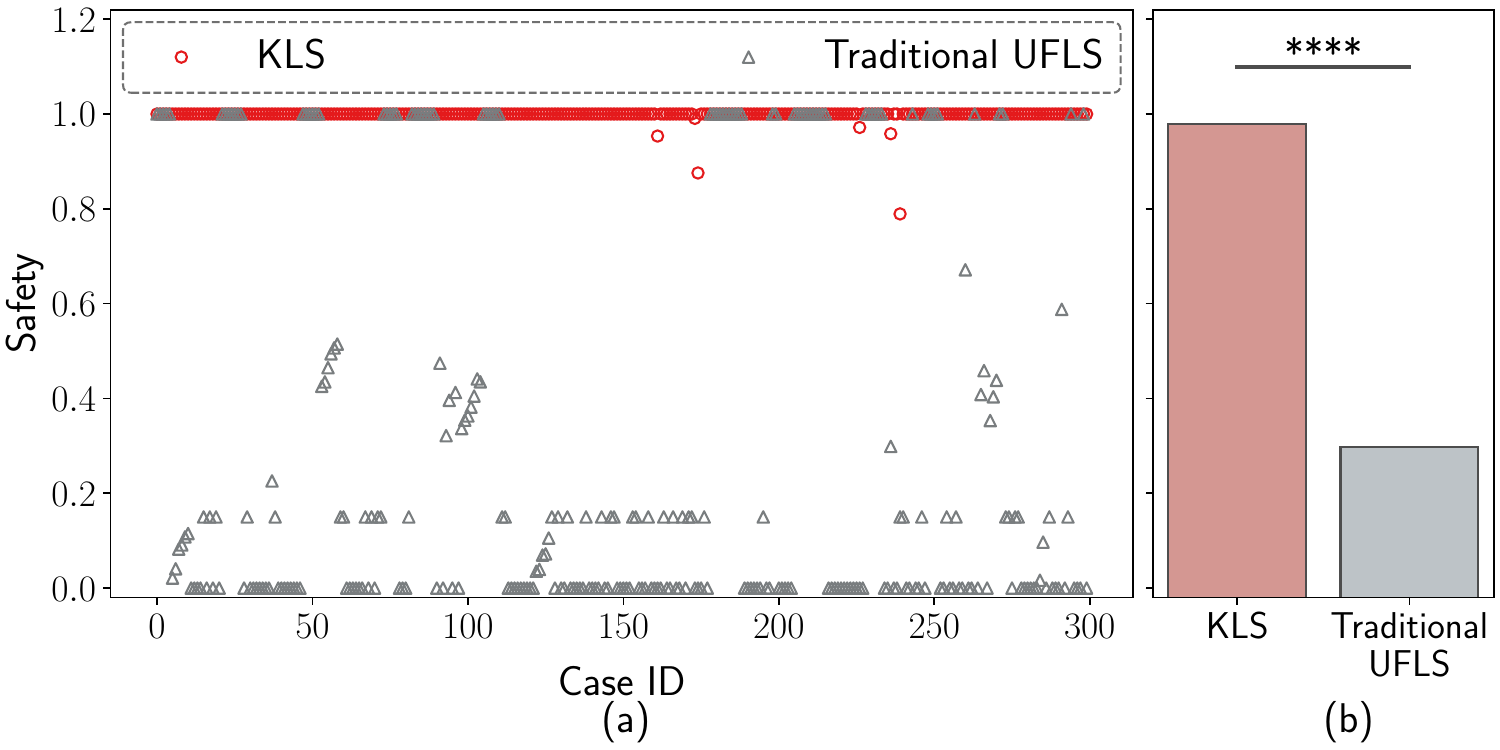}
    \caption{Comparison between the conventional UFLS and KLS.}\label{TRADITIONAL_CONTROL_SAFETY}
\end{figure}

To determine the load shedding amount for the conventional UFLS, we set the generator system inertia to the value specified by the original manufacturer. Additionally, we introduce a Generator 4 tripping fault, which causes a 350 MW power imbalance, to trigger the system dynamics. Generator 4 tripping fault results in a minimum system frequency of 48.6 Hz. By employing the conventional load shedding approach, we conducted tests with different proportions of load shedding to determine an optimal value. This optimal value ensures that both the frequency nadir and steady-state value are maintained within the safety range, while minimizing the amount of load shedding. Consequently, this determined value is considered as the preassigned load shedding proportion when the frequency reaches a preset threshold. Subsequently, we evaluated the system's safety in the testing set by assessing whether the frequency, after shedding the preassigned proportion of load, remained within the safety range. These results were compared with the outcomes obtained through KLS, as depicted in Fig. \ref{TRADITIONAL_CONTROL_SAFETY}. It is evident that shedding pre-assigned loads alone cannot ensure frequency safety under various operating conditions and power imbalances. However, KLS demonstrates its capability to adapt to such intricate variations in the inertia and the power imbalance.

\section{Conclusion}
In this paper, a novel data-enabled predictive control algorithm KLS, which adapts to diverse operating conditions and under frequency events, is introduced to achieve the optimal one-shot load shedding for power system frequency safety. To address approximation inaccuracies and the restriction of load shedding to discrete values, a safety margin tuning scheme is incorporated within the framework of KLS. Simulation results demonstrate that KLS effectively captures latent variables strongly correlated with the system inertia and power imbalance within a 300 ms time window after a fault occurs. The algorithm exhibits high prediction accuracy on both the training and testing sets, indicating its generalizability beyond the training set. Furthermore, the proposed safety margin tuning scheme enhances the system's frequency safety.

\bibliographystyle{IEEEtran}
\bibliography{ref}
\appendices
\section{Network Architecture and Training} \label{section_A}
\subsection{Network architecture and loss function}
\label{Network Architecture and Training}
\begin{figure}[t] 
    \centering
    \setlength{\abovecaptionskip}{-0.05cm}   
    \setlength{\belowcaptionskip}{-2cm}   
    \includegraphics[width=8.5cm]{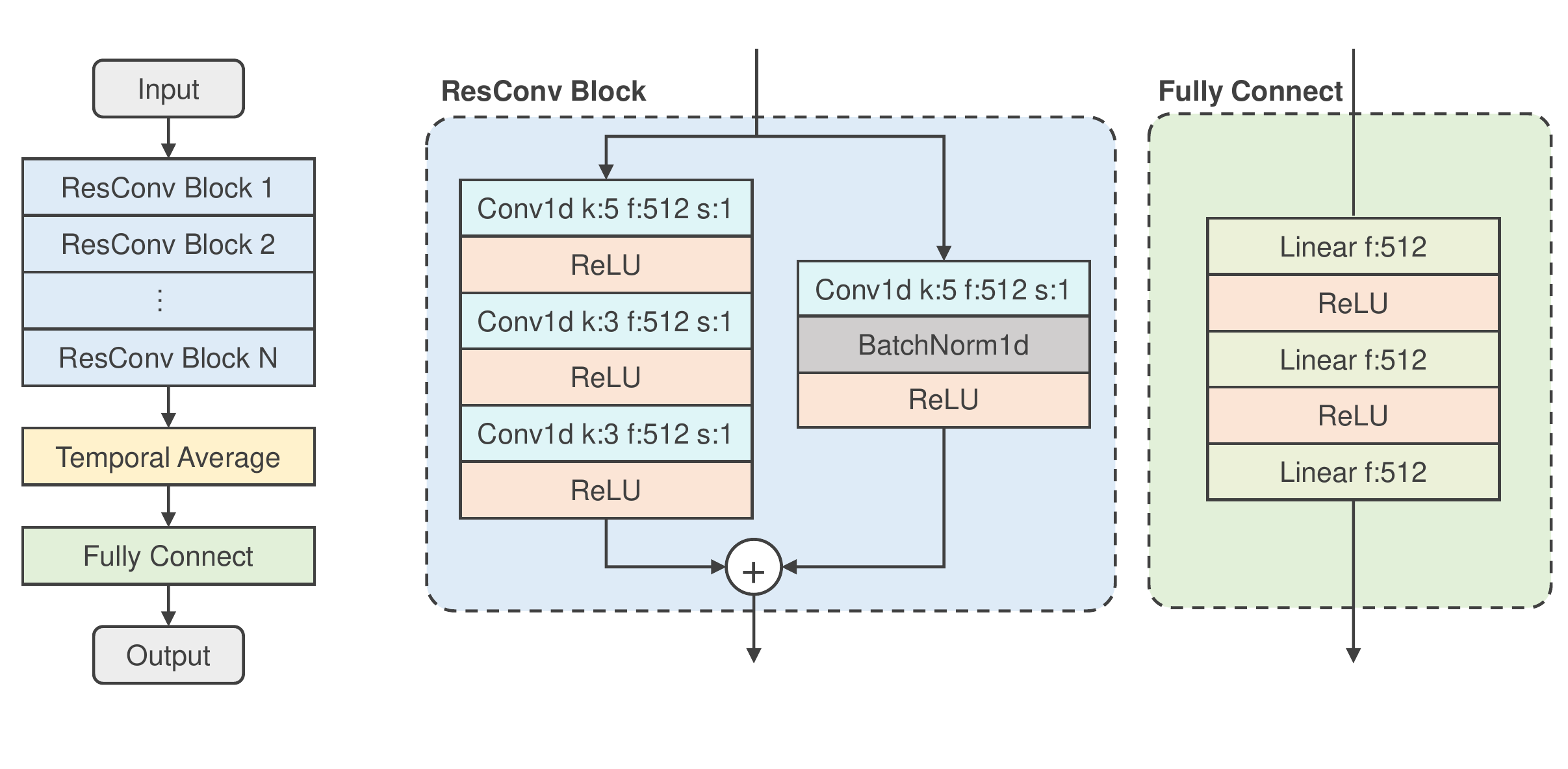}
    \caption{Network architecture of $\boldsymbol{\varphi}$.}\label{NN}
\end{figure}

The network architecture is depicted in Fig. \ref{NN}. The network is composed of three modules: the ResConv block, which acts as the latent extractor in this paper, as well as the Temporal Average and Fully Connected layers. The ResConv block utilizes convolutional layers to learn and identify the temporal patterns in the time series data. The inclusion of residual connections plays a vital role in enabling the stacking of multiple convolutional layers to effectively extract parameter variations in a state-space model from the time series data. The Temporal Average and Fully Connected layers are responsible for weighted averaging of time series features and feature extraction, respectively. 

The ResConv block comprises three convolutional layers with kernel sizes of 5, 3, and 3, respectively. A residual connection is incorporated by using a convolutional layer followed by a batch normalization layer. The Fully Connected module consists of three fully connected networks, employing ReLU activation functions.

The latent extractor extracts uncertain model parameters from the time-delay embedding of partially measured information. Temporal Average and Fully Connection layers take the concatenated vector of the extracted $\boldsymbol{m}$ with observable variables $\boldsymbol{x}$ and $\boldsymbol{y}$ as input and produces $\boldsymbol{\varphi} ({\omega }_{t-\tau :t},\boldsymbol{y}_{t-\tau :t})$ as output. The loss function described in Eq.\eqref{loss_function} ensures that $\boldsymbol{g}_{t}$ satisfies the requirements of global linearization. 

\begin{figure}[t] 
    \centering
    \setlength{\abovecaptionskip}{-0.05cm}   
    \setlength{\belowcaptionskip}{-2cm}   
    \includegraphics[width=10cm]{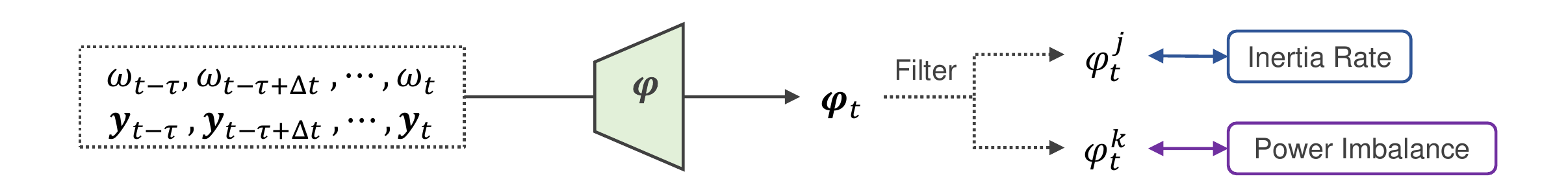}
    \caption{The latent extractor.}\label{LATENT_EXTRACTOR}
\end{figure}

\begin{figure}[t] 
    \centering
    \setlength{\abovecaptionskip}{-0.05cm}   
    \setlength{\belowcaptionskip}{-2cm}   
    \includegraphics[width=8.5cm]{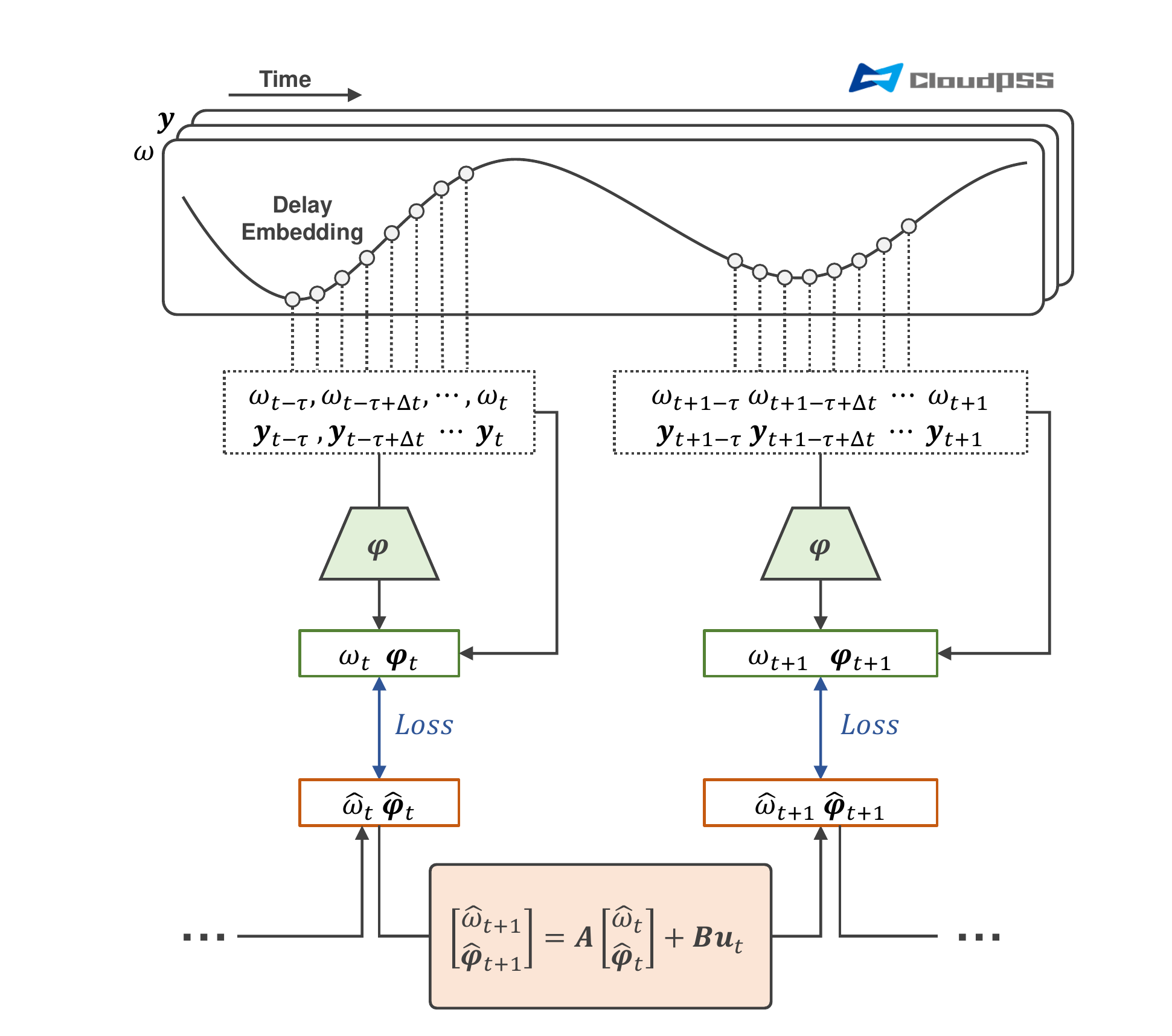}
    \caption{Data-driven KLS framework.}\label{FRAMEWORK}
\end{figure}

Fig. \ref{LATENT_EXTRACTOR} illustrates how to utilize the output vector of the latent extractor to obtain information strongly correlated with the variations in the system parameter. The "Filter" represents the selection of output variables in the latent extractor that are most relevant to the system inertia and power imbalance. The correlation coefficients between the selected variables, the system system inertia and power imbalance are calculated. The results are given in Fig. \ref{SCATTER_LATENTS_INRETIA_FAULT}. In general, our data-driven KLS framework is shown in Fig. \ref{FRAMEWORK}.

The loss function for the training of $\boldsymbol{\varphi}$ is given as follows.

\begin{align}
\mathcal{L}=\sum_{s=1}^{T-1}\sum_{t=s+1}^{T}\|\hat{\boldsymbol{g}}_{s\to t}-\boldsymbol{g}_{t}\|_2^2 \label{loss_function}
\end{align}

For each data point in the training set, $\hat{\boldsymbol{g}}_{s\to t}$ represents the predicted value at time $t$ in the Koopman linear system  with $\boldsymbol{g}_{s}$ as initial value. The loss function measures the discrepancy between the $T-1$ sequences of predicted values $\{\hat{\boldsymbol{g}}_{s\to t}, t=s+1,s+2,\dots,T\}$, and the true values at each time step $\{\boldsymbol{g}_{t}, t=s+1,s+2,...,T\}$. We use the $L_2$ norm to quantify the discrepancy between the predicted and true sequences.

\begin{algorithm}
  \SetKwData{Left}{left}\SetKwData{This}{this}\SetKwData{Up}{up}
  \SetKwFunction{Union}{Union}\SetKwFunction{FindCompress}{FindCompress}
  \SetKwInOut{Input}{input}\SetKwInOut{Output}{output}

  \Input{Training set $\{\omega_{t},\boldsymbol{u}_{t},\omega_{t-\tau:t},\boldsymbol{y}_{t-\tau:t} | t=1,2,\dots,T\}$}
  \Output{Optimized parameters of $\boldsymbol{\varphi}$, $\boldsymbol{A}$, and $\boldsymbol{B}$}
  \BlankLine
Random initialize the parameters of $\boldsymbol{\varphi}$, $\boldsymbol{A}$, and $\boldsymbol{B}$\;
Compute $\{\boldsymbol{g}_{t}| t=1,2,\dots,T\}$ according to Eq.\eqref{time_delay}\;
\While{not converge}{
  \tcp{Predict $\{\hat{\boldsymbol{g}}_{s\to t}| s=1,2,\dots,T-1;t=s+1,s+2,\dots,T\}$}
  \tcp{$\hat{\boldsymbol{g}}_{s\to t}$ is the predicted value at time $t$ with $\boldsymbol{g}_{s}$ as initial value}
  \For{$s\leftarrow 1$ \KwTo $T-1$}{
    $\hat{\boldsymbol{g}}_{s\to s+1}=\boldsymbol{A}\boldsymbol{g}_{s}+\boldsymbol{B}\boldsymbol{u}_{s}$\;
      \For{$t\leftarrow s+2$ \KwTo $T$}{
        $\hat{\boldsymbol{g}}_{s\to t}=\boldsymbol{A}\hat{\boldsymbol{g}}_{s\to t}+\boldsymbol{B}\boldsymbol{u}_{t-1}$\;
      }
  }
   Compute the prediction loss $\mathcal{L}$ according to Eq.\eqref{loss_function}\;
   Compute the gradient of $\mathcal{L}$\;
   Update the parameters of $\boldsymbol{\varphi}$, $\boldsymbol{A}$, and $\boldsymbol{B}$\;
}
  \caption{Procedure for training KLS}\label{algo_KLS}
\end{algorithm}\DecMargin{1em}

\end{document}